\documentclass[twocolumn,
preprintnumbers,
amsmath,amsfonts,amssymb,notitlepage,showpacs,aps,prx,
 superscriptaddress
 ]{revtex4-2}

\usepackage{xcolor}

\date{\today}

\usepackage{graphicx}
\usepackage{hyperref}
\usepackage[utf8]{inputenc}
\usepackage{tikz}
\usepackage{afterpage}

\tikzstyle{tensor}=[rectangle,draw=blue!50,fill=blue!20,thick]

\tikzstyle{H}=[rectangle,draw=red!50,fill=red!20,thick]

\tikzstyle{operator}=
[circle,draw=red!50,fill=red!20,thick]

\usepackage[normalem]{ulem}

\begin{document}
\title{Checkerboard Bose Hubbard Ladders using Transmon Arrays
}
\author{Pranjal Praneel}
\email{pp475@cornell.edu}
\affiliation{Laboratory of Atomic and Solid State Physics, Cornell University, Ithaca, New York}
\affiliation{Google
Quantum AI, Santa Barbara, California 93111}
\author{Thomas G. Kiely}
\email{thomaskiely@ucsb.edu}
\affiliation{Kavli Institute for Theoretical Physics, University of California, Santa Barbara, California, USA}
\author{Andre G Petukhov}
\email{apetukhov@google.com}
\affiliation{Google
Quantum AI, Santa Barbara, California 93111}
\author{Erich J Mueller}
\email{em256@cornell.edu}
\affiliation{Laboratory of Atomic and Solid State Physics, Cornell University, Ithaca, New York}

\begin{abstract} 
Adding a sublattice bias to the two dimensional Bose Hubbard model greatly enriches the available physics, and introduces knobs which can be used to control and interrogate the quantum state.  We describe the physics of this checkerboard Bose Hubbard model and how it can be explored using transmon arrays.  We show that the sublattice bias  brings the commensurate superfluid phase into an experimentally accessible regime,  and gives new probes. We  characterize the superfluid and insulating phases, with careful attention to finite size effects. 
\end{abstract}
\maketitle

\section{Introduction}






The defining problem of solid state physics is understanding the collective quantum mechanical behavior of particles in a periodic potential.  The 
phenomena
become 
rich when the interparticle interactions become strong, and the motions are strongly correlated with one-another.  The checkerboard Bose-Hubbard model is the ideal setting for exploring this physics.  We describe the behavior of this model and give a roadmap for  implementing it using current 
transmon based quantum processors \cite{
GoogleQuantumAI_Willow2025}.

It has long been recognized that the networks of superconducting elements (transmons) used to build many of today's quantum computers can  be viewed as controlable implementations of the 
Bose-Hubbard model \cite{Yanay2020,PRXQuantum.3.040314,Heras2015,PhysRevA.104.042602,orell2019,shi2023probing,Saxberg2022,Karamlou2024,Roushan2017,Ma2019,PhysRevResearch.4.013148,PhysRevLett.125.170503,PhysRevResearch.3.033043,PhysRevLett.126.180503,Yan2019,du2024probing,PhysRevLett.115.240501,li2024mapping,Gong2021,Kounalakis2018,roberts2023manybody,ticea2026observationdisorderinducedsuperfluidity}.  The quantum levels of each transmon are interpreted as the number of particles on each site.  Capacitive couplings let these ``particles" hop, while nonlinearities are equivalent to on-site interactions.  These circuits, however, operate in a regime where the coupling between the transmons is weak compared to the nonlinearities.  This limits our access to important parts of the phase diagram.  
 Here we propose introducing a sublattice  bias to overcome this difficulty.  This {\em checkerboard} model show additional physics beyond the vanilla Bose-Hubbard model, including topological charge pumping \cite{hayward2018,Citro2023,thouless,padhan2024,athanasio2024}.  We discuss state preparation, probes, and interesting feature of the phase diagram in ladders.

The insulating states in this model are particularly rich.
As was established in the 1990's, insulators are generally characterized by three quantities \cite{Ortiz1994,RestaBook,Resta1998}.  First, there is the energy gap $\Delta$, reflecting the minimum energy needed to create an excitation.
Second, the particles in an insulator are localized, allowing us to define a localization length $\xi$.  Third, one can characterize the positions of the localized particles with the polarization $\vec p$, which in a charged system would encode the dipole moment density.   
In a band insulator of non-interacting fermions, these quantities characterize the bands.  There, the gap is just the difference between the highest energy occupied state and the lowest energy empty state, 
the localization length is the size of the maximally localized Wannier state, and the polarization vector corresponds to the displacement of the center of mass of this Wannier state from the center of a unit cell.  The latter is typically defined in terms of a Berry phase.  Here we are working with an interacting system of bosons, so each of these quantities are true many-body observables.

In this context we are concerned about two separate classes of questions:  (1) What are the properties of the system in the thermodynamic limit, and (2) for an actual finite size experiment, how do you measure them?  We explore the phase diagram of the checkerboard system and characterize the quantum ground state.  We calculate the observables which could be used to probe this physics.  

We envision an experiment where one first produces the ground state of the system deep in the insulating state.  This is product state, and easy to engineer.  We then imagine slowly turning on the coupling between neighboring sites, adiabatically evolving the state.  One then measures relevant quantities, such as the  localization length.  One can thereby explore the superfluid-insulator transition, and study the quantum properties of the states. We model this state preparation, calculating the time required, and how it scales with system size.
Related considerations were discussed in \cite{PhysRevLett.126.045301,Gardas2017,Gericke2007,bakkalihassani2026revealingpseudofermionizationchiralbinding,Ebadi2021}, in slightly different contexts.  A number of other state preparation protocols could also be considered \cite{stateprep,Ma2019,Harrington2022,Saxberg2022,sequential}.


We focus on ladder geometries, as these are most approachable both in experiments and numerics.  They allow one to  interpolate between one dimensional (1D) and two dimensional (2D) systems by changing the number of legs.  Systems with fewer legs are easier to implement, but observing the superfluid-insulator transition in those settings requires large couplings between the transmons.  Additionally,  ladders possess some unique physical properties:  There is a distinction between ladders with even and odd numbers of legs and, as we will discuss, the insulating phase of the two-leg ladder displays a remarkable robustness \cite{crepin2011}.

\section{Model}\label{sec:BHM}
Transmon based quantum computers very naturally realize the Bose-Hubbard model \cite{Yanay2020,PRXQuantum.3.040314,Heras2015,PhysRevA.104.042602,orell2019,shi2023probing,Saxberg2022,Karamlou2024,Roushan2017,Ma2019,PhysRevResearch.4.013148,PhysRevLett.125.170503,PhysRevResearch.3.033043,PhysRevLett.126.180503,Yan2019,du2024probing,PhysRevLett.115.240501,li2024mapping,Gong2021,Kounalakis2018,roberts2023manybody,Saxberg2022}. A transmon can be viewed as an anharmonic quantum oscillator, whose Hilbert space basis states $|n\rangle$ are labeled by an integer $n$.  Unlike a harmonic oscillator, where the energy is linear in $n$, the energy here is a non-linear function, 
$E_n=(\hbar \omega)\times(n+1/2)+(\hbar\eta/2)\times n(n-1)+\cdots$,
where the neglected terms are higher order in $n$.  
If one interprets $n$ as the number of {\em particles} on a lattice site, then this non-linearity becomes a short-range interaction:  There is an energetic difference between the state with two ``particles" on the same site as compared to when they are on different sites.  
Operators $\hat a$ and $\hat a^\dagger$ connect the states in this Hilbert space: $\hat a |n\rangle=\sqrt{n} |n-1\rangle$, $\hat a^\dagger |n\rangle=\sqrt{(n+1)}|n\rangle.$ Capacitively coupling the transmons allows these excitations to move around.  Number non-conserving processes are off-resonant, and hence can be ignored \cite{ticea2026observationdisorderinducedsuperfluidity}. 

 After transforming into a rotating frame and neglecting off-resonant terms, the Hamiltonian for a transmon array is
\begin{equation}\label{eq:transmon_bhm}
    \frac{\mathcal{H}}{\hbar}
    =\sum_{\langle i,j\rangle\in\Lambda}g_{ij}(\hat a^\dagger_i \hat a_j+{\rm H.c.})+\sum_{i
    }\frac{\eta}{2}\hat n_i(\hat n_i-1)+\frac{\delta_i}{2} \hat n_i.
\end{equation}
Here $i$ indexes the transmons, $\Lambda$ is a set of unordered pairs $\langle i,j\rangle$ which designate the connectivity of the network, and $\hbar=h/(2\pi)$ is Planck's constant.  We report all quantities on the right hand side in units of Hz, though only the ratios matter for determining the physical state.  We envision a lattice where 
$\langle i,j\rangle\in\Lambda$ denotes nearest neighbors. 
The operators $\hat a_i,\hat a_i^\dagger,$ and $\hat n_i$ are raising, lowering and number operators for the $i$'th transmon.  The term $g_{ij}$ encodes coupling between connected transmons.  
Flux tuned couplers can allow $g_{ij}$ to be changed dynamically \cite{Yan2018,ibmcouplers}.
The detuning $\delta_i=\omega_i-\bar{\omega}$ represents the oscillation frequency of the $i$'th transmon, measured relative to some reference frequency $\bar{\omega}$, which was introduced when we tranformed to the rotating frame \cite{ticea2026observationdisorderinducedsuperfluidity}. 
We take $\delta_i=\delta$ on one sublattice, and $\delta_i=-\delta$ on the other.
Given that $N=\sum_i n_i$ is conserved by this Hamiltonian, the system's properties are independent of $\bar{f}$.   The non-linearity, $\eta<0$ is  fixed.  We take an average of one particle per site, and consider $g_{ij}/2\pi\lesssim 30$ MHz,  $\eta/2\pi\sim250$ MHz and $\delta/2\pi\sim250$ MHz.  

Equation~\ref{eq:transmon_bhm} is
the checkerboard Bose-Hubbard model if we identify $t=-\hbar g_{ij}$,  $U=\hbar \eta$. 
Note that under this mapping, the Hubbard interaction $U$ is strictly negative.  Nonetheless we can explore the physics of the positive $U$ Bose Hubbard model.  For a bipartite lattice, the spectrum of the repulsive and attractive models are simply related.  Each repulsive many-body state $|\nu\rangle$ with energy $E_\nu$ is gauge-equivalent to an attractive state with energy $-E_\nu$.  

We envision an experiment where one begins in the ground state of the repulsive model for vanishing $g$.  This is a product state which is easily created.  It corresponds to the highest energy state of the attractive model.  We then imagine slowly changing Hamiltonian parameters.  By the adiabatic theorem we will remain in the ground state of the repulsive model (highest energy state of the attractive model) as  the parameters are varied. Sec.~\ref{sec:stateprep} explores the timescales for this process.

\begin{figure*}
    \centering
    \includegraphics[width=\textwidth]{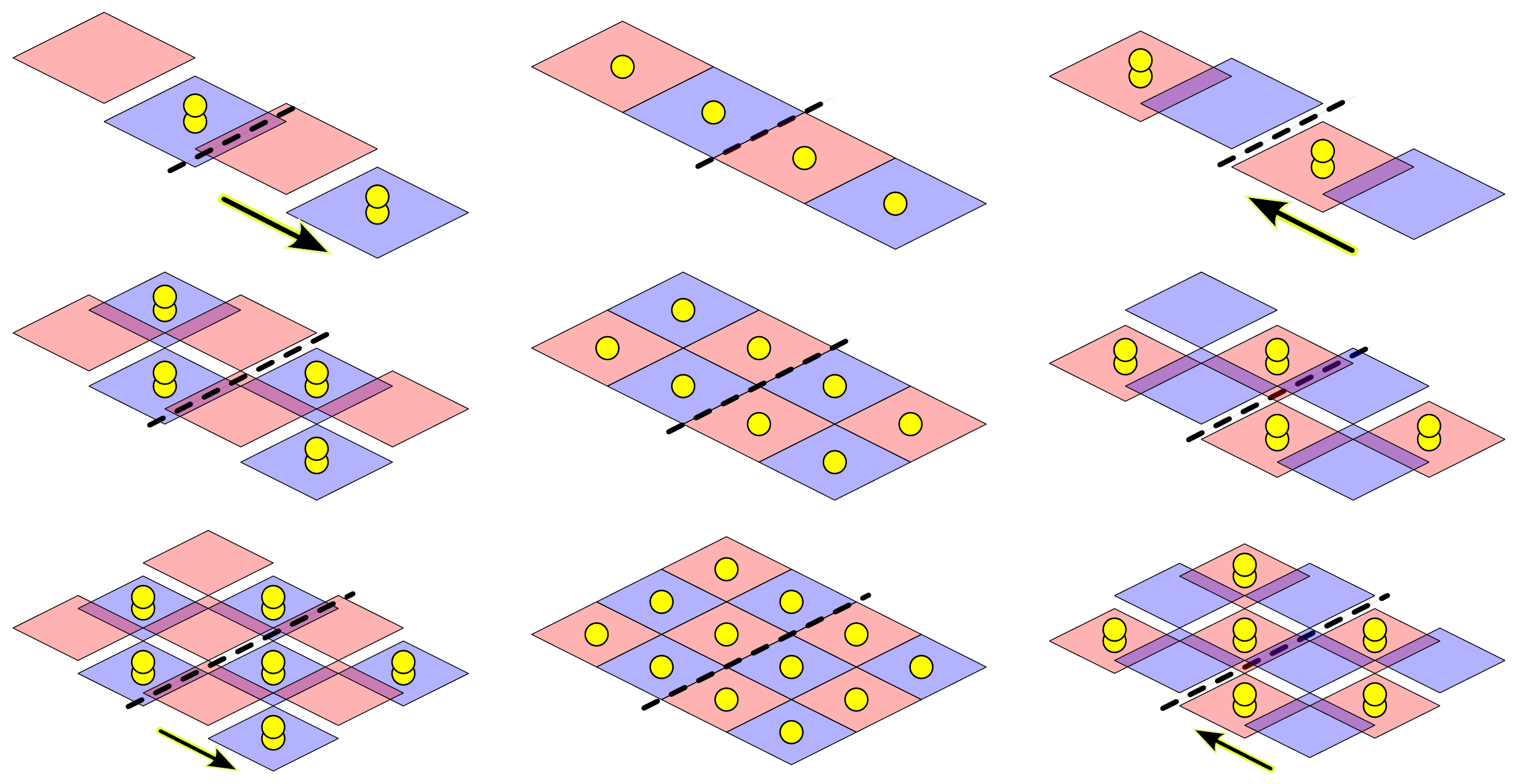}
    \caption{Length 4 segments of a checkerboard ladders of width $N=1$ (top), $N=2$ (middle), and $N=3$ (bottom).  From left to right the sublattice bias $\delta$ is tuned from a negative to positive value, visualized by the relative heights of the red and blue sites.  When the sites are uncoupled, $g\ll\eta$, the particles are all localized on individual sites.  The unit filled ($\bar{n} = 1$) configurations are shown in this atomic limit.  A dashed line breaks the segment into two unit cells.  An arrow represents the polarization vector. There is no arrow displayed when the polarization $p=0$.  
    } 
    \label{fig:oddeven}
\end{figure*}

We will consider ladder geometries, consisting of $N$ legs, and $L$ rungs (see Fig.~\ref{fig:oddeven}).  These are readily implemented in experiments, and let one interpolate between one and two dimensions.  We will largely focus on the case where there is one particle per site.

\subsection{Mapping to spin models}
\label{hc}
Transmons are often used as qubits in the regime where occupations are restricted to $n=0,1$. As reviewed below, in this {\em qubit regime} they are 2-level systems which map on to the hardcore Bose-Hubbard model or the XX spin model \cite{Yanay2020}.    We can introduce a similar mapping in the checkerboard case, where, at appropriate fillings, the occupations in one sublattice are restricted to $n_A=1,2$, and the other to $n_B=0,1$.  Note, our analysis is not restricted to this limit -- we just introduce it to give an easily understood baseline.

As a first step  we derive the spin model for the more familiar case where the coupling between qubits $g_{ij}$, and the sublattice bias, $\delta$ are small compared to the non-linearity $g,\delta\ll\eta$. 
In the standard Bose-Hubbard variables, this corresponds to $t\ll U$.
 Under these circumstances is energetically unfavorable to ever have more than one particle per site, $n_i=0,1$.  
We  assign spins as $|0\rangle\to|\!\downarrow\rangle, |1\rangle\to|\!\uparrow\rangle$.  Truncated to these states, and neglecting an unimportant constant, Eq.~(\ref{eq:transmon_bhm}) reduces to the $XY$ model,
\begin{equation}\label{eq:transmon_XY}
\frac{\mathcal{H}_{\rm eff}}{\hbar}=\sum_{\langle i,j\rangle\in \Lambda}
\frac{g_{ij}}{2}(\sigma^x_i\sigma^x_j+\sigma^y_i\sigma^y_j) +\sum_i \frac{\delta_i}{4} \sigma^z_i
\end{equation}
where $\sigma^\nu$ are Pauli matrices for $\nu=x,y,z$.  The numerical factors come from the fact that $(\sigma^x_i\sigma^x_j+\sigma^y_i\sigma^y_j)\to 2 (a_i^\dagger a_j+a_j^\dagger a_i)$ and $\sigma_i^z\to 2n_i-1$.
Note that $\eta$ itself does not appear in this effective model.  It has been replaced by a constraint. 
An important feature of this hard-core model is that in one dimension it can be mapped onto a free fermion problem through a Jordan Wigner transformation.  Thus this  is an exactly solvable limit.
 
We now generalize this construction, and consider the case where $\delta\sim \eta \gg g$.
In that case a doubly-occupied site on the low-energy A-sublattice is nearly degenerate with a singly-occupied site on the B-sublattice.  One can then restrict the occupations to $n_A=1,2$ and $n_B=0,1$.
 We again denote these two possibilities as $|\downarrow\rangle,|\uparrow\rangle$, and can write Hamiltonian as a XY model.  Neglecting the constant offset,
\begin{align}
\frac{\mathcal{H}_{\rm eff}}{\hbar}&=\sum_{\langle i,j\rangle\in \Lambda}
J_{ij}(\sigma^x_i\sigma^x_j+\sigma^y_i\sigma^y_j) +\sum_{i} (-1)^i \frac{s}{4} \sigma^z_i 
\nonumber
\\\label{xy}
J_{ij}&=\frac{g_{ij}\xi_i\xi_j}{2} \qquad s = (\delta-\eta)/2
\end{align}
Here $\xi_i=1$ if $i\in A$ and $\xi_i=\sqrt{2}$ if $i\in B$.  For the checkerboard case this implies that $J_{ij}=g_{ij}/\sqrt{2}$.  Again in 1D this can be mapped onto noninteracting fermions. 

As already explained, at times we will use this mapping to gain extra insight into the physics of Eq.~(\ref{eq:transmon_bhm}).  Most of our calculations, however, work with the full model.  





\section{Insulating States}\label{sec:insulators}

At commensurate densities and small inter-site coupling, 
$g \ll\eta$ ($t\ll U$ in standard Hubbard model language), the ground state of Eq.~(\ref{eq:transmon_bhm}) is an {\em insulator}.  This means that it does not support bulk particle currents.

Consider, for example, the case where there is on average one particle per site, $\bar n=1$ and vanishing sublattice bias, $\delta=0$.  As depicted in the center column of Fig.~\ref{fig:oddeven}, as $g\to 0$, one has a product state with exactly one particle per site.  Turning on a small coupling, $g$, leads to local number fluctuations, where a particle hops to a neighboring site, creating an empty site and a doubly occupied site.  The insulating state does not admit bulk mass currents, and these holes and doublons remain close to one-another. We will define a localization length, $\xi$, which quantifies the average separation.

As illustrated by the first and third column of Fig.~\ref{fig:oddeven}, if one makes the sublattice bias large enough, the nature of the $g=0$ state changes.  The lowest energy configuration becomes a checkerboard of doubly occupied sites.  At finite $g$, one would imagine that one can continuously tune between the uniform and checkerboard configurations by delocalizing the atoms between sites.  Remarkably, such a continuous deformation is only possible if there are an even number of legs, or one breaks reflection symmetry across a lattice site. 

This feature is related to a symmetry protected 
$\mathbb{Z}_2$ topological invariant that can be extracted from the x-component of the polarization vector $\pi_x$~\cite{Ortiz1994,RestaBook}.  
The polarization is the vector sum $\vec{\pi}=\sum_{\alpha=1}^\nu ({\vec R}_\alpha-{\vec R}_\alpha^{(0)})$, where $\vec{R}_\alpha$ are the locations of the $\nu$ particles that are within one unit cell, and $\vec{R}_\alpha^{(0)}$ is the center of the unit cell.  If one assigns a charge $e$ to the particle, and places an appropriate opposite charge at the center of the unit cell, $\vec{\pi}$ is proportional to the electric dipole moment per unit cell.  Depending on context, one might divide $\vec{\pi}$ by the number of particles in a unit cell, or by the volume of the unit cell, giving either the polarization per particle or the polarization density.  In our case the relevant quantity will be the polarization per unit length along the long axis of the ladder, and hence we define $p=\pi_x/\ell_x=\pi_x/2$, where $\ell_x=2$ is the length of the unit cell of the imposed checkerboard potential.



As emphasized in \cite{RestaBook}, the polarization is not uniquely defined by the positions of the particles, as one needs to decide which particles are associated with which unit cells.  In practice one makes this assignation by considering adiabatically connecting the state to a reference.  Regardless, it is natural to only define $\vec\pi$ up to a lattice vector that connects the center of two unit cells.  Consequently we only define $p$ modulo $1$.  

For an $N$ leg checkerboard ladder at unit filling, each unit cell contains $2N$ sites and $\nu=2N$ particles. In Fig.~\ref{fig:oddeven} we 
illustrate the $g\to0$ configurations and their polarization vectors.
For an even number of legs, the polarization vector is zero in all cases.
For odd numbers of legs, the situation is different.  The three  configurations shown all have different polarization vectors.  Though it is worth noting that the checkerboard configurations in the first and third columns have polarizations that are equivalent up to a multiple of the unit cell size.


Our model is symmetric under reflection across a lattice site, which takes $p\to-p$.  The configuration is only invariant under this operation if  $p=0,1/2$ (modulo 1), defining two distinct topological classes.  
Thus, unless we break the symmetry, $p$ cannot be continuously changed -- and there must be a phase transition separating states with different $p$.  This feature is discussed in more detail in Sec.~\ref{pumping}.  




For a finite length ladder with open boundary conditions,  edge effects make the unit cells at different locations inequivalent. Under those circumstances, it is natural to  
average over all unit cells, similarly to Ref. \cite{Carrasquilla2013} defining
\begin{align}\nonumber
\hat P &= \frac{1}{L_x}\sum_{\alpha=1}^{N_p} (\hat{x}_\alpha- x_\alpha^{(0)})\\
&= 
\frac{1}{L_x}\sum_j x_j (\hat n_j-\bar n).
\label{pdef}
\end{align}
Here $N_p$ is the total number of particles, $L_x$ is the number of rungs in the ladder, and $\bar n=1$ is the average density.  The quantities $\hat P, \hat x_\alpha$, and $\hat n_j$ are all operators.  


In 
Ref.~\cite{Resta1998}, the polarization was formalized by considering a length $L_x$ system with periodic boundary conditions.  It is then natural to introduce the twist operator
\begin{equation}\label{eq:w_nleg}
    \hat W=\exp\left(\frac{2\pi i}{L_x}\sum_{j=1}^Lx_j(\hat n_j-\bar n)\right),
\end{equation}
which, as the complex exponential of $\hat P$, is its periodic generalization.
Within this formalism, the polarization in a state $|\psi\rangle$ is defined as $2\pi p={\rm Arg}\langle \psi|\hat W|\psi\rangle$.  
Consistent with our previous discussion,  this expression only defines $p$ modulo 1. This construction of $p$ reduces to our original definition $p=\pi_x/2$ in the atomic limit where every particle is in a well defined location, and $\pi_x$ is well defined. 
Note that $\hat W$ can be interpreted as a momentum-space translation operator, and $p$ is related to the Berry phase under moving through a non-contractable path in momentum space, $\theta_B=2\pi p$ \cite{Ortiz1994}. 

For some of our calculations we work in the thermodynamic limit, $L_x\to\infty$.  There it is convenient to define a generalized twist operator,
\begin{equation}\label{twist}
    \hat W_k=\exp\left(ik\sum_{j}x_j(\hat n_j-\bar n)\right).
\end{equation}
In Sec.~\ref{sec:Thermodynamic_limit} we make an infinite matrix product state (iMPS) ansatz, and show that $\langle \hat W_k \rangle =w_k^{L_x}$, where $L_x=\infty$ is the length of the ladder.  The parameter $w_k$ can be numerically calculated from the iMPS wavefunction.
The polarization is then
$p=\lim_{k\to 0} {\rm Arg}(w_k/k)$.

We now define the localization length 
in terms of the average fluctuation in the position of each particle
\begin{align}
\xi^2&=\left\langle\frac{1}{N_p}\sum_\alpha(\hat x_\alpha^2- \bar x_\alpha^2)\right\rangle\\
&=L_x\left(\langle \hat P^2\rangle-
\langle \hat P\rangle^2\right).
\end{align}
Here $\bar x_\alpha=\langle \hat x_\alpha\rangle$.
This localization length can  be interpreted as a measure of the fluctuations of the center of mass of the particles, scaled by the particle number.  One can formally relate the localization length to the twist operator through a cumulant expansion,
$\langle e^A\rangle = \exp(\langle A\rangle+(\langle A^2\rangle-\langle A\rangle^2)/2+\cdots)$.  This yields
\begin{eqnarray}\label{wk}
w_k&=& \left(\left\langle e^{ik L_x \hat P}\right\rangle\right)^{1/L_x}
= \exp\left(
ik\langle \hat P\rangle - \frac{1}{2} k^2  \xi^2\right).
\end{eqnarray}
In particular, the polarization can be extracted from the phase $\theta_B\equiv \arg(w_k)/k$.
As already explained, the thermodynamic limit of  $w_k$ can be extracted from our infinite matrix product state calculations.

\subsection{Connecting the topological sectors: Charge Pumping}\label{pumping}
As detailed by Hayward et al. \cite{hayward2018} in the context of cold atoms, the physics from the previous section can be most dramatically studied via a charge pumping experiment \cite{Citro2023,thouless}.  Similar ideas were also proposed by Padhan et al. \cite{padhan2024}, with a focus on two-leg ladders.  Some considerations about charge pumping in Josephson Junction arrays was given in \cite{athanasio2024}.

To investigate this physics, one generalizes Eq.~(\ref{eq:transmon_bhm}) to the Rice-Mele-Bose Hubbard model by alternating the strength of the the hopping between sites.  In one dimension, where the site locations are indexed by integers $j$ one takes 
\begin{equation}\label{km}
g_{ij}=\delta_{j,i+1}(-t+(-1)^i \lambda).
\end{equation}
The non-uniform hopping breaks the reflection symmetry, and $p$ will be a continuous function of the sublattice bias $\delta$ and the hopping asymmetry $\lambda$.

An important feature is that any closed path in the $\delta-\lambda$ parameter space must bring $p$ back to its original value, up to an integer.  If that integer is nonzero, one has transported a quantized amount of charge \cite{thouless}.  In an annular geometry, an experiment could directly measure this topological charge pumping by reading out the state of each transmon and directly measuring the expectation value of the twist operator, $\langle W\rangle$.  One should be able to follow the motion of charge, and demonstrate the quantization of the charge pumping.  For the Rice-Mele Bose Hubbard model the amount of charge pumped is simply proportional to the number of times that one winds around the origin in parameter space \cite{hayward2018}.

In multi-leg ladders there are more choices in how to break the symmetry.  Some of these will lead to charge pumping, while others simply bring the charges back to their original location.  For even number of legs, the configurations in Fig.~\ref{fig:oddeven} are all in the same topological sector.  This feature {\em does not preclude  charge pumping}. 
Even in multi-leg ladders, symmetry breaking terms, such as Eq.~(\ref{km})  allow $p$ to continuously vary, and there are  closed paths in parameter space which increase $p$ by an integer.  The special feature of the even leg ladders is that one can connect the states in Fig.~\ref{fig:oddeven} without breaking reflection symmetry.    The polarization $p$ is fixed on paths which do not break reflection symmetry, 
and therefore they
rearrange the atomic positions without leading to any net charge transport.

\section{Phase Diagram}{\label{sec:Thermodynamic_limit}}
\label{sec:imps}

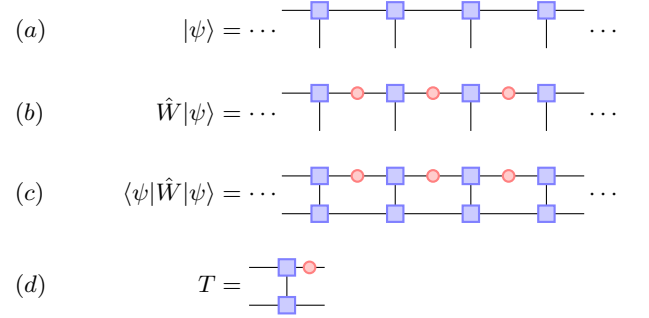
\begin{figure}
\begin{align*}
   &(a)&\qquad |\psi\rangle&=\cdots\raisebox{-0.25\height}{\begin{tikzpicture}
    \draw (0.5,1) -- (4.5,1);
    \draw (1,1) -- (1,0.5);
    \draw (2,1) -- (2,0.5);
    \draw (3,1) -- (3,0.5);
    \draw (4,1) -- (4,0.5);
    \node[tensor] at (1,1) {};
    \node[tensor] at (2,1) {};
    \node[tensor] at (3,1) {};
    \node[tensor] at (4,1) {};
\end{tikzpicture}}\cdots
\\[1em]
&(b)&\qquad \hat W|\psi\rangle &=\cdots\raisebox{-0.25\height}
    {\begin{tikzpicture}
    \draw (0.5,1) -- (4.5,1);
    \draw (1,1) -- (1,0.5);
    \draw (2,1) -- (2,0.5);
    \draw (3,1) -- (3,0.5);
    \draw (4,1) -- (4,0.5);
    \node[tensor] at (1,1) {};
    \node[operator,scale=0.5] at (1.5,1) {};
    \node[tensor] at (2,1) {};
    \node[operator,scale=0.5] at (2.5,1) {};
    \node[tensor] at (3,1) {};
    \node[operator,scale=0.5] at (3.5,1) {};
    \node[tensor] at (4,1) {};
    \end{tikzpicture}}\cdots
    \\[1em]
    &(c)&
    \qquad \langle\psi|\hat W|\psi\rangle &=
\cdots\raisebox{-0.4\height}{\begin{tikzpicture}
    \draw (0.5,1) -- (4.5,1);
    \draw (0.5,0.5) -- (4.5,0.5);
    \draw (1,1) -- (1,0.5);
    \draw (2,1) -- (2,0.5);
    \draw (3,1) -- (3,0.5);
    \draw (4,1) -- (4,0.5);
    \node[tensor] at (1,1) {};
    \node[tensor] at (1,0.5) {};
    \node[operator,scale=0.5] at (1.5,1) {};
    \node[tensor] at (2,1) {};
    \node[tensor] at (2,0.5) {};
    \node[operator,scale=0.5] at (2.5,1) {};
    \node[tensor] at (3,1) {};
    \node[tensor] at (3,0.5) {};
    \node[operator,scale=0.5] at (3.5,1) {};
    \node[tensor] at (4,1) {};
    \node[tensor] at (4,0.5) {};
\end{tikzpicture}}\cdots\\[1em]
&(d)&
    \qquad T &=
\raisebox{-0.4\height}{\begin{tikzpicture}
    \draw (0.5,1) -- (1.5,1);
    \draw (0.5,0.5) -- (1.5,0.5);
    \draw (1,1) -- (1,0.5);
    \node[tensor] at (1,1) {};
    \node[tensor] at (1,0.5) {};
    \node[operator,scale=0.5] at (1.3,1) {};
\end{tikzpicture}}
\end{align*}
\caption{Graphical representations of the infinite matrix product states used to calculate the phase diagram.  (a) Each blue box represents the wavefunction on one rung of the ladder -- the vertical line represents the physical degrees of freedom, while the horizontal lines represent internal variables $\{s_i\}$, summed over in Eq.~(\ref{mps}).  (b) The twist operator $\hat W$ acting on $|\psi\rangle$ can be expressed as  a matrix where a unitary transformation is placed on the internal bonds.  (c) The expectation of the twist operator is expressed as the contraction of a matrix product state, and can be calculated from the largest eigenvalue of the mixed transfer matrix. (d) Mixed transfer matrix used to calculate the polarization and localization length.
}\label{fig:mps}
\end{figure}

\subsection{Methods}\label{mpsmethods}

Our iMPS ansatz is equivalent to writing
\begin{equation}\label{mps}
|\psi\rangle = \sum_{\{s_j\}} \cdots|\phi_{s_1s_2}\rangle_1 |\phi_{s_2s_3}\rangle_2\cdots
\end{equation}
where $|\phi_{st}\rangle_i$ is a wavefunction for the $i$'th 
unit cell, which consists of two rungs of the $N$-leg ladder, and hence contains $2N$ sites.  These states in each unit cell are indexed by two integers,
$s,t=1,2,\cdots \chi$.  Thus $|\phi_{st}\rangle_i$ can be interpreted as a matrix of quantum states.  Our ansatz is translationally invariant and we take  $|\phi_{st}\rangle_i$ to be identical to $|\phi_{st}\rangle_j$.  In our numerics each $|\phi_{st}\rangle$ is  written as a finite matrix product state over the sites in one unit cell, though for the purposes of this presentation we will treat it as a singular entity.  Figure~\ref{fig:mps} (a) shows a graphical representation of the Eq.~(\ref{mps}).  Each box represents a term, $|\phi_{s_j\,s_{j+1}}\rangle_i$.  The lines connecting the boxes correspond to the indices $s_j$ which are summed over.  The lines extending from the bottom of each box represents the physical Hilbert space  within each unit cell
\cite{SCHOLLWOCKreview}.

To identify the various phases we calculate the twist operator from Eq.~(\ref{twist}), which can be expressed as a sum over the rungs 
\begin{equation}\label{twist2}
    \hat W_k=\exp\left(ik\sum_{i}x_i(\hat n_i^{(r)}-\bar n^{(r)})\right).
\end{equation}
Here $\hat n_i^{(r)}$ is the number of particles on the rung labeled $i$.  It is the sum, $\hat n_i^{(r)}=\sum_j \hat n_j$ of number operators on each site $j$, on that rung.

To calculate the twist operator, we use an approach that is related to one given by Zaletel et al. \cite{Zaletel2014}.  We first define $\hat N_i^R=\sum_{j=i}^\infty (\hat n_j^{(r)}-\bar n^{(r)})$ as the excess number of particles to the right of the $i$'th rung.  We then note that $
\hat P=L_x^{-1}\sum_j x_j (\hat n_j^{(r)}-\bar n^{(r)})=L_x^{-1}\sum_j\left(x_j N_j^R-x_j N_{j+1}^R\right).$
We then shift the index on the second term and use $x_{j}-x_{j-1}=1$ to get
$\hat P= L_x^{-1} \sum_j \hat N_j^R$.  This allows us to write the twist operator in a translationally invariant form:
%
%
%
\begin{eqnarray}
    \hat W
    &=&\prod_{i}\exp\left(ik \hat N^R_i\right).
\end{eqnarray}
Since the total number of particles are conserved in this system, we can choose the $|\phi_{s_i s_{i+1}}\rangle_i$ so that there are a definite number of particles 
{ in that unit cell.}  The states to the right of {cell} 
$i$ are indexed by $s_{i+1}$,
\begin{equation}
|\psi^{(R)}_{s_{i+1}}\rangle
= \sum_{s_{i+2},s_{i+3},\cdots} 
|\phi_{s_{i+1}s_{i+2}}\rangle_{i+1} |\phi_{s_{i+2}s_{i+3}}\rangle_{i+2}\cdots
\end{equation}
The state $|\psi^{(R)}_{s_{i+1}}\rangle$ with fixed $s_{i+1}$ must have a well defined number of particles, 
$\hat N^R |\psi^{(R)}_{s}\rangle= (N^R)_{s} |\psi^{(R)}_{s}\rangle$.
That is, each value of the bond index $s$ is associated with a number $(N^R)_s$, which physically corresponds to the number of particles to the right of that bond.
Thus applying $\hat W$ to $|\psi\rangle$ is equivalent to multiplying each term in Eq.~(\ref{mps}) by the product
\begin{equation}
W_{\{s\}}=\prod_i e^{i k N^R_{s_i}}.
\end{equation}
Graphically, this corresponds to the diagram in Fig.~\ref{fig:mps} (b), where we include a diagonal matrix on each bond. 

The expectation value  $\langle W\rangle$ then involves contracting the tensor network in Fig.~\ref{fig:mps} (c).  This contraction is found by constructing the mixed transfer matrix
\begin{equation}
T_{st;s^\prime t^\prime}=
\langle \phi_{st}|\phi_{s^\prime t^\prime}\rangle e^{i k N^R_{s^\prime}},
\end{equation}
{ which is shown graphically in Fig.~\ref{fig:mps} (d).}
The largest eigenvalue of this matrix is $w_k$.
For an insulating state, the iMPS wavefunction can be used to construct the ground state on a ring of size $L_x$~\cite{Zaletel2014}; in this case, one would find $\langle W\rangle = w_k^{L_x}$.  The polarization and correlation length are extracted from Eq.~(\ref{wk}).  

In addition to calculating the twist operator, we also estimate the many-body excitation gap $\Delta$, which corresponds to the minimum energy needed to excite the system.  The insulating state has a finite $\Delta$, while for the superfluid $\Delta=0$.  To calculate this gap, we look at the spectrum 
$\epsilon_n^{(\chi)}$
of the `local Hamiltonian', which is constructed during the VUMPS procedure.  One can interpret $\epsilon_1^{(\chi)}$ as the smallest energy of an excitation that is orthogonal to $|\psi\rangle$ and formed by changing only 
a single one of the $|\phi_{st}\rangle$ in Eq.~(\ref{mps}).  
One expects that
$\lim_{\chi\to\infty} \Delta_\chi$ is a good approximation to $\Delta$ as long as the matrix product state is sufficiently expressive \cite{chepiga2017,chepiga2020,eberharter2023}.  
In Appendix~\ref{gapextrapolation} we describe our extrapolation procedure and provide some additional physical reasoning about the excitation spectrum.



We find the superfluid-insulator boundaries by searching for where $\Delta=0$,  We verify that these locations also correspond to  $\xi=\infty$, or equivalently $w_k=0$.  The superfluid-insulator transition is in the XY-universality class, and can also be found by looking at the behavior of the Luttinger parameter.  


\begin{figure}
    \centering
    \includegraphics[width=\columnwidth]{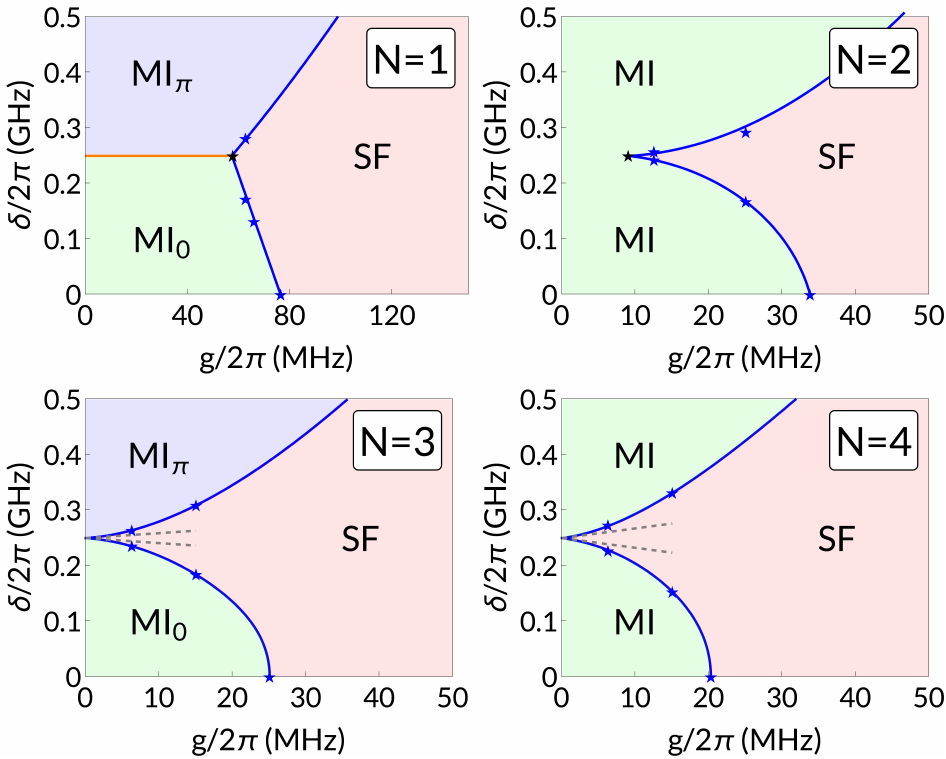}
    \caption{Phase diagrams of the 
    Bose-Hubbard model on an $N$-leg ladder with a checkerboard potential (Eq.~(\ref{eq:transmon_bhm}) at total filling $\bar n=1$, as calculated from an infinite matrix product state ansatz. 
    Stars correspond to the locations of the phase boundaries found in our iMPS calculations, while solid lines represent interpolations or extrapolations.
    For the plots shown, the anisotropy $\eta/2\pi=0.25$~GHz. All ladder geometries host superfluid (SF) and insulating (MI) phases. Odd-leg ladders show two topologically-distinct insulating phases, ${\rm MI}_{\theta_B}$, with Berry phases $\theta_B=0,\pi$. This leads to a critical line (orange) in the $N=1$ phase diagram below $g/2\pi\approx57$~MHz, where the extended superfluid phase is unstable.     
    Even-leg ladders differ in that the insulators are topologically indistinguishable. For $N=2$, the SF phase terminates at $g/2\pi\sim9$~MHz, below which a gap opens up. For $N\geq 3$, the extended superfluid phase is stable as $g\to0$ for $|\delta-\eta|\leq cg$ where $c\sim\mathcal{O}(1)$ is a constant. This behavior is shown by the gray dashed lines.}
    \label{fig:phasediagram}
\end{figure}

\subsection{Results}
\label{infiniteresults}
Figure~\ref{fig:phasediagram} shows the phase diagram of infinitely long $N$-leg ladders, 
with one particle per site.  In these figures we take $\eta/2\pi=0.25$ GHz -- changing this frequency simply rescales each axis.  For odd N there are 3 phases: the superfluid and two toplogically distinct insulating states, written as MI$_0$ and MI$_\pi$, corresponding to $p=0,1/2$, or $\Theta_B={\rm Arg}(w_k)/k=0,\pi$.  Cartoons of these insulating states, depicting $g\to 0$ atomic configurations, are shown in Fig.~\ref{fig:oddeven}.
As anticipated in Sec.~\ref{sec:insulators}, there is only one type of insulator for even-N ladders, though for 4 or more legs the insulating phase is broken into two disjoint regions by a superfluid intrusion.  

Practically speaking, the most significant aspect of these diagrams is that adjusting $\delta$ enables a significant reduction in the critical coupling $g_c$ needed to generate a superfluid, particularly when $N>1$. Given experimental constraints of $g/2\pi \lesssim 30$MHz, it is extremely difficult to produce a superfluid in an $N=1$ system.  This challenge is overcome by increasing $N$ and operating at finite $\delta$.

There are a number of other interesting features in Fig.~\ref{fig:phasediagram}.  First, the $N=1$ and $N=2$ cases are qualitatively different from the higher $N$ ladders.  When $N=1$ the smallest coupling that hosts a superfluid phase is $g/\eta\approx0.23$ and occurs when $\delta=\eta$.  For smaller $g$, there is a direct, { continuous} transition between the two topologically distinct insulating phases.  Exactly on the critical line, $\delta=\eta$, the low energy properties are described by a gapless Luttinger liquid with $K<2$, for which the sublattice bias is a relevant perturbation, driving an instability towards either the MI$_0$ or MI$_\pi$ phase \cite{giamarchi}. 
For the 2-leg ladder, the smallest coupling that hosts a superfluid phase is $g/\eta\approx0.04$.  At smaller $g$ one can continuously tune between the insulators in Fig.~\ref{fig:oddeven} simply by changing $\delta$.  There is no phase transition, and the sublattice population simply increases with $\delta$.

For all $N>2$, the phase diagrams look similar. { The only qualitative distinction is the difference in the topological invariants of the two Mott lobes on odd-leg ladders.} At $\delta=\eta$ the superfluid phase extends to arbitrarily small $g$.  As shown by dashed lines in the figures, the phase boundaries come in with a finite slope: $|\delta_c-\eta| \propto g$.  



\begin{figure}
    \centering
    \includegraphics[width=\columnwidth]{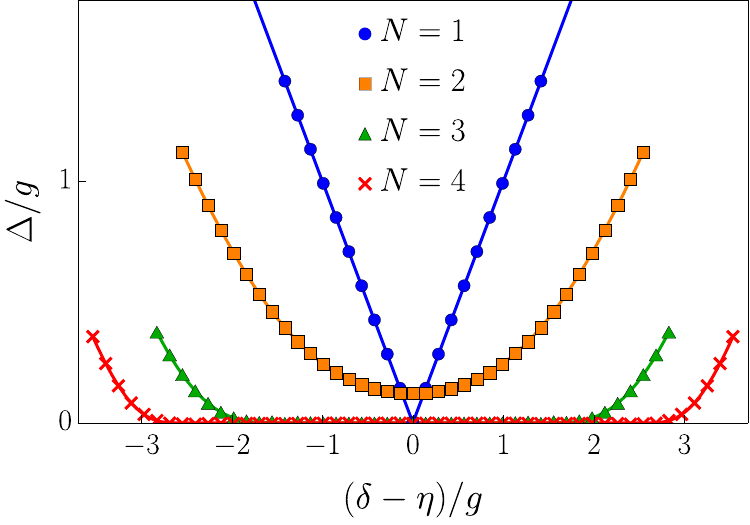}
    \caption{Ground state gap $\Delta$ of Eq.~(\ref{xy}) at 
    zero magnetization, corresponding to the small $g$ limit of Eq.~(\ref{eq:transmon_bhm}) at unit filling with sublattice bias $\delta \sim \eta$.
    Colored dots show the extrapolated iMPS numerical results (error bars are smaller than marker size) while solid lines are fits, where appropriate (see main text). When $N=1$, the model maps onto non-interacting fermions and may be solved exactly, with $\Delta=|\delta-\eta|$. When $N=2$, system remains gapped even at $\delta=0$. For $N\geq 3$, the gap closes for some finite region around $\delta=0$, which is qualitatively consistent with two-dimensional ($N\to\infty$) limit of Eq.~(\ref{xy}).
    }
    \label{fig:nleg_gaps}
\end{figure}

\subsection{Excitation gap in the hard core limit}\label{hardcore}

The most distinctive feature in Fig.~\ref{fig:phasediagram} is the behavior of the system for small $g\neq0$ near $\delta\approx\eta$.  For $N=1$ there is a direct transition between insulating phases.  For $N=2$ there is no phase transition.  For $N\geq 3$ one sees two transitions as one varies $\delta$.

As anticipated by the discussion in Sec.~\ref{hc}, we can model  this limit by mapping onto a spin model.  When $g$ is small, and $\delta\approx \eta$, the lower energy sublattice will contain either $1$ or $2$ particles, while the higher energy sublattice contains $0$ or $1$ particles.  We denote the states as $|\!\!\downarrow\rangle$ and $|\!\!\uparrow\rangle$, and the Hamiltonian reduces to the XX-model in Eq.~(\ref{xy}).  The condition that there is one boson per site in our original model corresponds to a vanishing net magnetization.  This XX model can also be interpreted as a {\em hard-core} Bose Hubbard model, but with only one particle for every 2 sites.

In 1D ($N=1$) this spin model is exactly solvable, as it can be mapped onto non-interacting fermions via a Jordan-Wigner transformation \cite{JordanWigner,ColemanMBP2004}.  It generically corresponds to a band insulator, with a gap to neutral excitations of $\Delta=|\delta-\eta|$.  At $\delta=\eta$ one has a 1D metal of non-interacting fermions -- i.e. a gapless Luttinger liquid with Luttinger parameter $K=1$.

For larger $N$ the model is no longer exactly solvable, but we can numerically calculate its properties using  iMPS techniques. 
The resulting energy gaps are shown in Fig.~\ref{fig:nleg_gaps}.  For $N=1$ one sees a ``V" shaped gap which vanishes at a single point.  For $N=2$ the gap is always finite, and forms a ``U".  For $N\geq 3$ there is a finite range of $\delta/g$ over which the gap vanishes.  

The gapped behavior of the two-leg ladder is very counterintuitive.  It implies that the hard core Bose Hubbard model on a 2-leg ladder is an insulator at density 1/2.    This unexpected result was studied in \cite{crepin2011}.  It can be understood by separately considering individual rungs of the ladder.  There are two single particle states:  one which is reflection-symmetric, and the other which is antisymmetric.  
{ The ground state can be understood as a Mott insulator 
of the symmetric modes.}
This picture can be verified by making the hopping between the legs of the ladder, $g_\perp$, stronger than the hopping 
{between rungs of the ladder}, $g_\parallel$.  If $g_\perp\gg g_\parallel$, the splitting between the symmetric and antisymmetric modes is enhanced, stabilizing the insulator.  Conversely, if $g_\perp\ll g_\parallel$ the gap closes.   In \cite{ying2014}, Ying et al. considered $N$ leg ladders, arguing that by making $g_\perp$ sufficiently small one could find insulators with densities 
{$\bar n=m/N$ with $m\in\mathbb{Z}$.}  When $g_\parallel=g_\perp$, however, none of these insulators are found in the hard core limit, except the case $m=1$ and $N=2$.

For 3 and 4-leg ladders, the gaps in Fig.~\ref{fig:nleg_gaps} appear to approach zero with vanishing slope.  Such a behavior is consistent with the $(1+1)$-dimensional XY (BKT) universality class, where
\cite{BKT1,BKT2,BKT3}
\begin{equation}\label{eq:bktgap}
    \Delta=\Delta^0 \exp \frac{a}{\sqrt{|\delta_c - \delta|}}.
\end{equation}
Here $\delta_c$ is the critical value of $\delta$, at which the gap vanishes, while $\Delta_0$ and $a$ are constants.  All of these quantities depend on the number of legs, $N$.  In Fig.~\ref{fig:nleg_gaps} we fit the gaps to Eq.~(\ref{eq:bktgap}), finding excellent agreement.  For $N=3$ we find
$\delta_c=\eta\pm 1.27(5) g$, and for $N=4$, $\delta_{c}=\eta\pm 2.47(9)g$.

This functional form of the gaps suggests that the low energy behavior of these ladders is the same as an effective one dimensional model.  This would suggest that the low energy physics is described by a conformal quantum field theory, which can be characterized by a conformal central charge, 
$C$, which counts the number of low energy modes \cite{szasz2020}.  In particular, a Luttinger liquid would display 
$C=1$,  
corresponding to having two low energy modes that can be identified as
``right-movers" and ``left-movers"~\cite{giamarchi}.

The conformal central charge can be extracted from a Matrix Product State wavefunction by the scaling of the bipartite entanglement entropy, $S$, with bond dimension, $\chi$~\cite{kiely2022,pollmann2009,szasz2020}.  We repeat our iMPS calculations with a series of bond dimensions ranging from 80 to 280. 
{We compute the entanglement entropy and the iMPS correlation length, $\xi_{\chi}=-1/\ln\lambda_{\chi}$, where $\lambda_\chi$ is the subdominant eigenvalue of the transfer matrix~\cite{kiely2022,pollmann2009}, and fit them to
\begin{eqnarray}
    S(\chi)&=&\frac{C}{6}\ln(\xi_{\chi}).
\end{eqnarray}
}
Indeed, we find that $C=1$ in the gapless phase of these ladders, implying that there {are only two low-energy modes.}
%
%
%
%
We understand this behavior
as phase-locking between the different legs, leaving just the gapless modes associated with the total phase and total density~\cite{crepin2011}.

\begin{figure}
    \centering
    \includegraphics[width=0.9\linewidth]{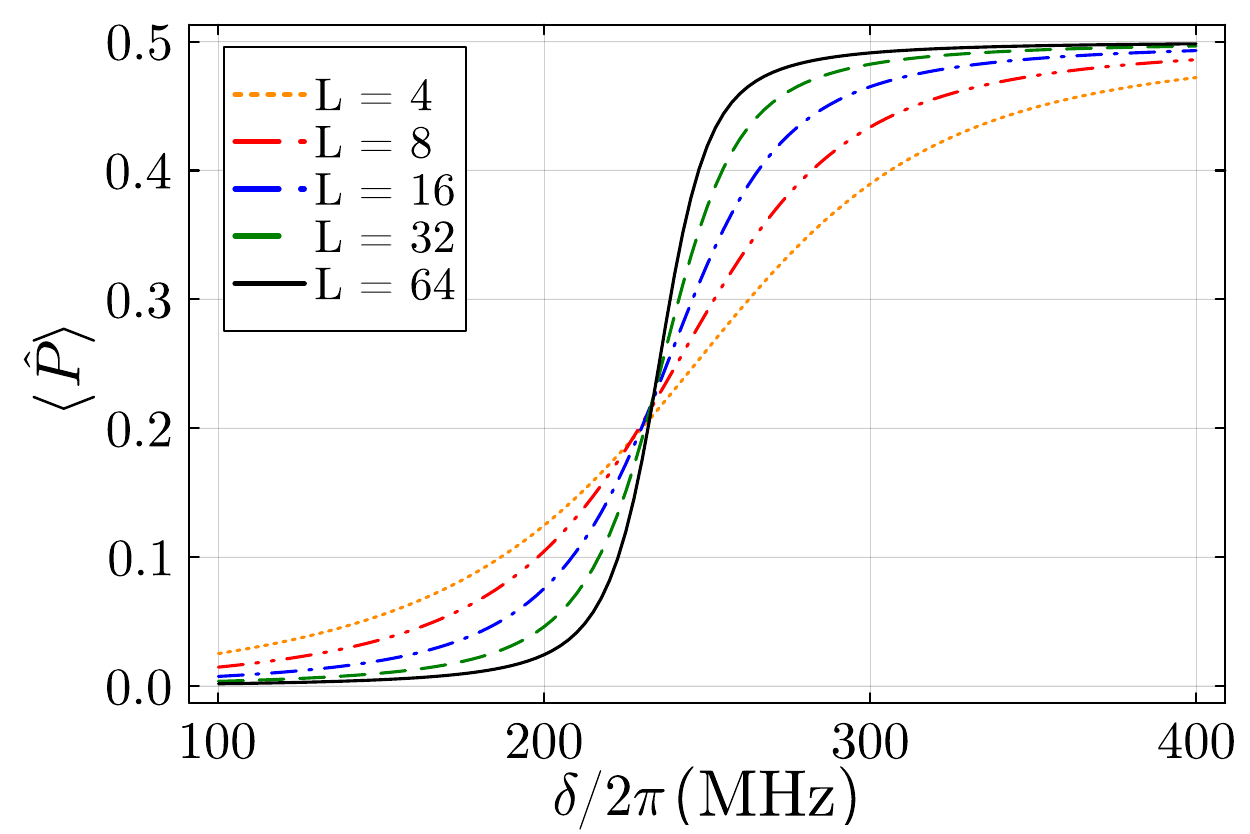}
    \caption{Expectation value of the polarization $\langle \hat P\rangle$ in a 1D Bose-Hubbard chain of Length $L$, with $\eta/2\pi=250$~MHz and $g/2\pi=40$~MHz.  As one changes the sublattice bias, $\delta$, the particles in each unit cell move from one lattice site to the other.  In the limit $L\to\infty$ the polarization  discontinuously jumps from $0$ to $1/2$ as $\delta$ crosses $\eta$.}
    \label{fig:Pol1D}
\end{figure}

\section{Finite Size Systems}\label{finite}
Any practical experiment would be performed on lattices with 10's or at best 100's of sites \cite{Roushan2017,Ma2019,PhysRevResearch.4.013148,PhysRevLett.125.170503,PhysRevResearch.3.033043,PhysRevLett.126.180503,Yan2019,du2024probing,PhysRevLett.115.240501,li2024mapping,Gong2021,Kounalakis2018,roberts2023manybody,Saxberg2022}. Thus, it is essential to understand how the features from Sec.~\ref{sec:insulators} and \ref{sec:Thermodynamic_limit} appear in such small systems.  A useful approach is to treat these experiments as analog simulations and perform `finite size scaling'.  One could repeat an experiment with multiple system sizes, and extrapolate the results to the thermodynamic limit.

In this finite size setting, the $x$ component of the polarization vector is related to the the center of mass.
The two topologically distinct insulators in odd leg ladders can be distinguished by $\langle \hat P\rangle$ where, as already introduced,
\begin{equation}
\hat P =\frac{1}{L_x}\sum_i x_i  (\hat n_i-\bar n).
\end{equation}
Here $i$ labels the individual sites of the lattice, $x_i$ is the $x$-coordinate of the lattice site, $\hat n_i$ is the number of particles and $\bar n=1$ is the average particle density.  One expects that $\langle \hat P\rangle\approx 0,\pm 1/2$ in the finite analog of the MI$_0$ and MI$_{\pi}$ phases.  In the finite size system, it is not quantized, as $\langle \hat P\rangle$ is sensitive to the structure of the edge.

To calculate the properties of these finite size systems we again make a variational matrix-product state ansatz.  We use the finite DMRG algorithm to optimize the wavefunction \cite{SCHOLLWOCKreview}, and increase the bond dimension $\chi$ until the results do not change.

In Fig.~\ref{fig:Pol1D} we show how the polarization varies with the sublattice bias $\delta$ in an $N=1$ leg system for fixed $g/2\pi=40$ MHz and $\eta/2\pi=250$ MHz.  As $L$ becomes larger the finite size effects become smaller, and $\langle \hat P\rangle$ approaches 0 for $\delta<\eta$.  Conversely $\langle \hat P\rangle\to 1/2$ as $L\to \infty$ for   $\delta>\eta$.

%

To distinguish the superfluid and insulating phases we look at the
fluctuations of the polarization $\langle \Delta\hat{P}^2 \rangle= \langle \hat{P}^2 \rangle-\langle \hat{P} \rangle^2$,
which serves as a measure of particle delocalization. As described in Sec.~\ref{sec:insulators}, in the Mott insulating phase we define the localization length $\xi$ in terms of the fluctuations of the polarization as 
$\langle \Delta\hat{P}^2 \rangle=\xi^2/L$.

In Appendix~\ref{sec:ll} we argue that in the 
superfluid phase we should expect that for large $L$, $\langle \Delta\hat{P}^2 \rangle$ will approach an $L$ independent constant, which is related to the Luttinger parameter $K$,
\begin{equation}\label{llp}
\begin{split}
    \lim_{L_x \to \infty} \left\langle \Delta\hat{P}^2 \right\rangle
    &= \frac{7\zeta(3)}{2\pi^4}K.
\end{split}
\end{equation}
Here $\zeta(3)=1.20206$ is Ap\'ery's constant.
Physically $K$ is related to the compressibility and the speed of sound, $u$, by $K=u\pi(\partial n/\partial\mu)$. Smaller $K$ corresponds to a stiffer fluid. Importantly, at the superfluid-insulator transition $K=2$.  This value is a universal feature of the Kosterlitz-Thouless transition.  

\begin{figure*}[tbph]    \includegraphics[width=\columnwidth]{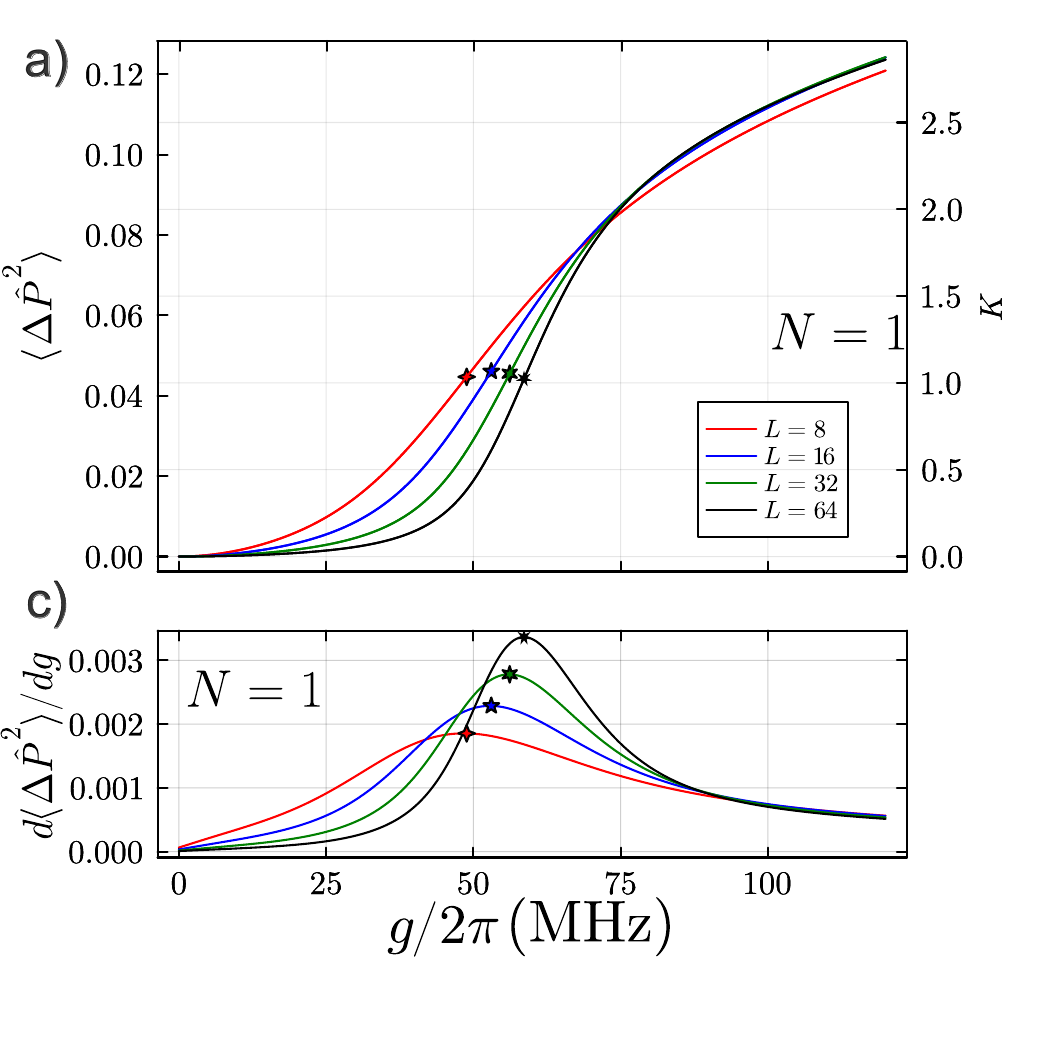}
    \includegraphics[width=\columnwidth]{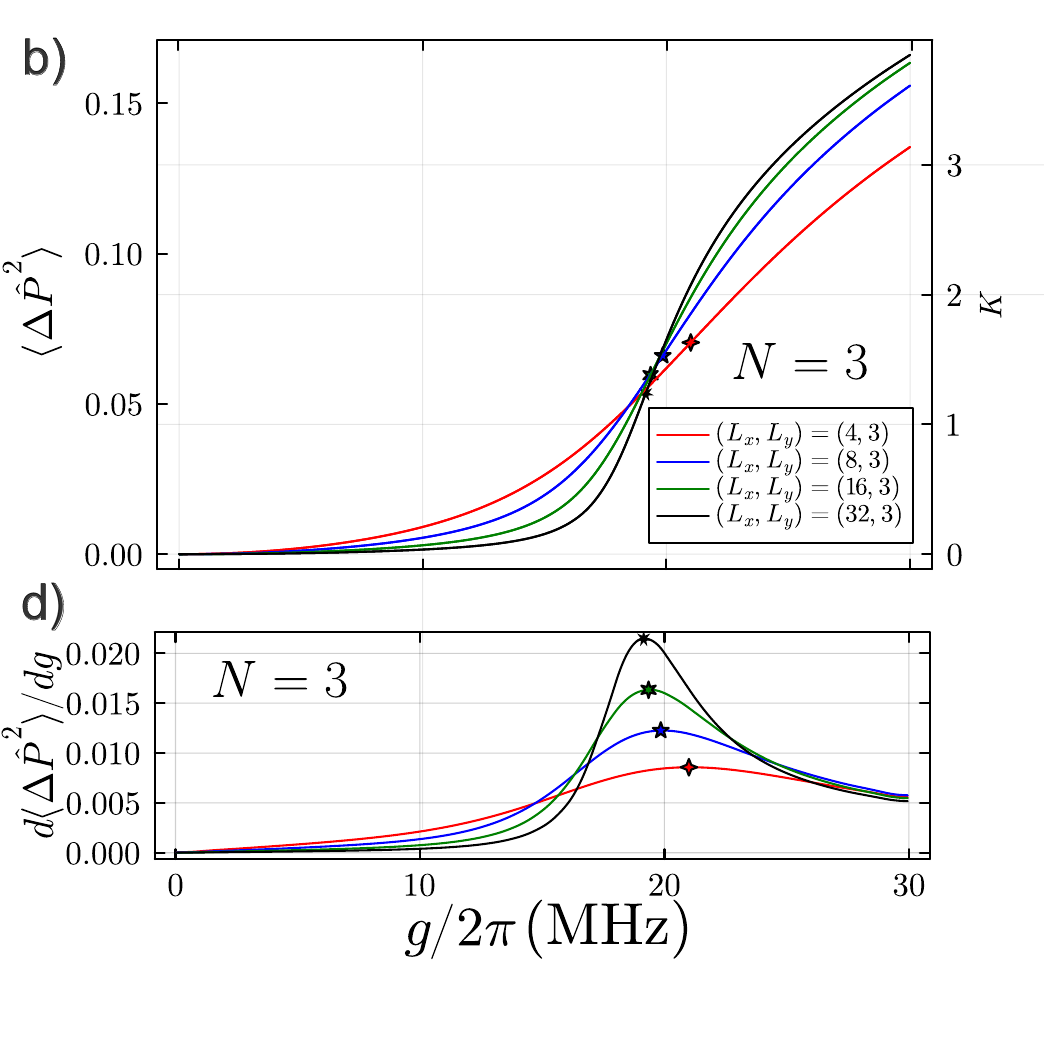}
    \caption{
    Left: (a) Fluctuations in the polarization and (b) its derivative as a function of $g$ for a 1D chain with $\eta/2\pi =$ 0.25~GHz and $\delta = \eta/2$.  The right hand axis uses Eq.~(\ref{llp}) to convert the fluctuations into a luttinger constant.  Symbols correspond to the inflection points where $d \langle \Delta \hat P^2\rangle/dg=0$. Right: Fluctuations in the polarization and its derivative as a function of $g$ for a N=3-leg ladder with $\eta/2\pi =$ 0.25GHz and $\delta = \eta/2$.  
    }
    \label{fig:XflucVariation}
\end{figure*}

Thus we expect that the superfluid and insulating phases can be distinguished by the $L$ dependence of $\langle \Delta \hat P^2\rangle$.  
In Fig.~\ref{fig:XflucVariation} we show 
$\langle \Delta \hat P^2\rangle$ for both $N=1$ (top-left) and $N=3$ legs (top-right) for a range of system sizes.  In both cases we have taken $\eta/2\pi=0.25$GHz and $\delta=\eta/2$.
We also show the slope $\partial \langle \Delta \hat P^2\rangle/\partial g$.
In each case there is a critical $g_c$, and for $g<g_c$ the polarization fluctuations fall towards zero with increasing system size, consistent with $\langle \Delta \hat P^2\rangle=\xi^2/L$.  Conversely, for $g>g_c$ one instead finds the fluctuations grow with system size, approaching an asymptotic value, consistent with the expectation in Eq.~(\ref{llp}).  The right hand scale of the figure shows the corresponding value of the Luttinger parameter $K$ from Eq.~(\ref{llp}).  

The maximum of the slope $\partial \langle \Delta \hat P^2\rangle/\partial g$ should occur at a value of $g$ which approaches $g_c$ in the thermodynamic limit, where the polarization fluctuations should display a discontinuous jump.  
In Fig.~\ref{fig:XflucVariation} we mark the positions of the maximum slope.  In Sec.~\ref{polextrapolation} we extrapolate to the $L=\infty$ limit, finding that $g_c/\eta=0.24,~0.076$ for the 1-leg and 3-leg ladders at $\delta=\eta/2$, respectively. Experiments can follow this same procedure to find the critical point.

The maximum value of 
$\partial \langle \Delta \hat P^2\rangle/\partial g$ grows with $L$, consistent with the expectation that
 in the thermodynamic limit,
the $ \langle \Delta \hat P^2\rangle$ will show a discontinuity at $g=g_c$, jumping from $0$ to $7\zeta(3)K_c/(2\pi^4)$, where $K_c=2$.  For a finite size system the steepest part of the curve will occur half-way between these values.  Thus as $L$ is increased we see the y-coordinates of the symbols in Fig.~\ref{fig:XflucVariation} (a,b) approach a value which corresponds to $K=1$.

\subsection{Finite size scaling of Polarization Fluctuations}\label{polextrapolation}

Here we analyze the data from Fig.~\ref{fig:XflucVariation} where we find the maximum of $d \langle \Delta \hat P^2\rangle/dg$ as a function of system size $L$.  We fit the system size dependence to the expected behavior of the
Kosterlitz-Thouless transition \cite{Harada1997, Zuo2021}. 

\begin{figure}
\includegraphics[width=\columnwidth]{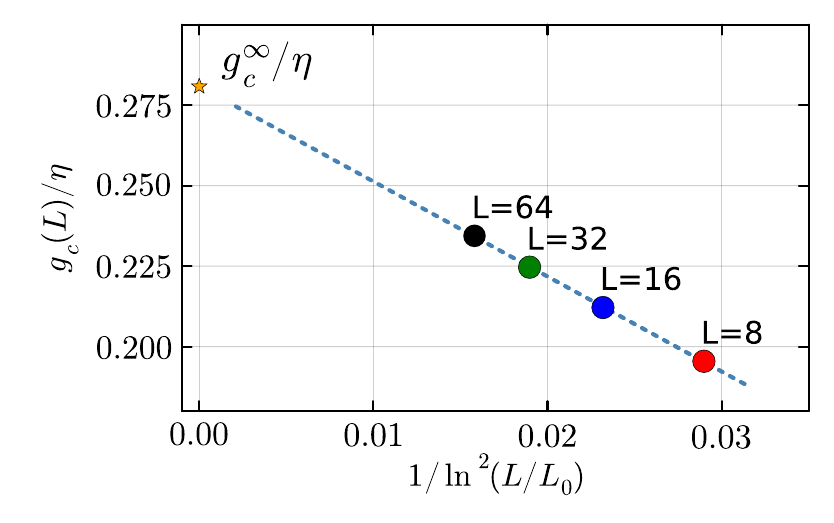}
    \caption{
    Finite size scaling for the critical coupling in a 1D chain.   Here $g_c(L)$ correspond to the marked points in Fig.~(\ref{fig:XflucVariation}) (a,c), where $d^2 \langle \Delta \hat P^2\rangle/dg^2=0$, for a chain with $N=1$, and sublattice bias $\delta=\eta/2$.  The scaling length $L_0=0.0225$ is chosen by fitting to Eq.~(\ref{eqn:ext_fit_bkt}), and the dashed line shows the predicted scaling relationship.  The extrapolation to $L=\infty$ should correspond to the critical point for the superfluid-insulator transition. }
    \label{fig:gc_1D_est}
\end{figure}
\begin{figure}
\includegraphics[width=\columnwidth]{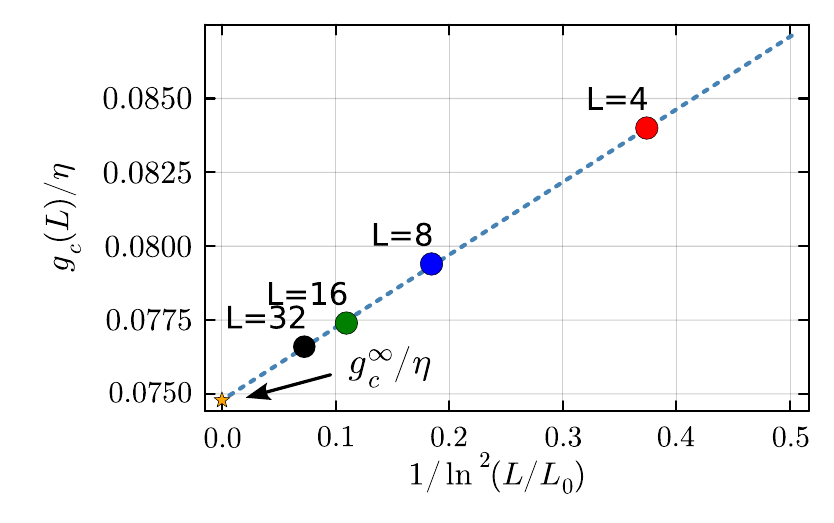}
    \caption{
    Finite size scaling for the critical coupling in a 3-leg ladder.
    Here the data is taken from Fig.~(\ref{fig:XflucVariation}) (b,d)
    and $L_0=0.7795$. }
    \label{fig:gc_3leg_est}
\end{figure}

The peak in the susceptability $\chi=d\langle \Delta \hat P^2\rangle/dg$ of the finite system occurs when the correlation length is of order the system size, $\xi\sim L$.  We define $g_c(L)$ as the coupling constant at this peak.  Near the critical point the correlation length scales as $\xi \propto \exp(a/\sqrt{g-g_c})$, where $a$ is a constant and $g_c=g_c^\infty$ is the critical point \cite{fisher1989}.
It then follows that,
\begin{align}\label{eqn:ext_fit_bkt}
    \frac{g_c(L)}{\eta} = \frac{g_c^\infty}{\eta} +\frac{A}{\ln^2\left(\frac{L}{L_0}\right)}, 
\end{align}
where $L_0,A$, and $g_c^{\infty}$ are constants, independent of system size.
Figures~\ref{fig:gc_1D_est} and \ref{fig:gc_3leg_est} use  this scaling relationship to determine the critical point from the behavior of the polarization.

For each system size in Fig.~(\ref{fig:XflucVariation}), we extract  $g_c(L)$ as the location of the inflection point of the polarization fluctuation curve. 
We then perform a non-linear least square fit to determine the parameters $g_c^\infty, A, L_0$ in Eq.~(\ref{eqn:ext_fit_bkt}).  Figures~\ref{fig:gc_1D_est} and \ref{fig:gc_3leg_est} show that the system size dependence appears to be well described the scaling relationship.   
The respective $L_0$ and $A$ for the 1-leg and 3-leg ladders are $L_0^{1D} \approx 0.0225$ and $A^{1D} \approx -2.9511$, and $L_0^{3leg} \approx 0.7795$ and $A^{3leg} \approx 0.0246$.
Considering the experimentally relevant nonlinearity of $\eta/2\pi = 250$ MHz, we find that the transition occurs 
 at $g_c^{1D}/2\pi \approx 70.21  $ MHz and $g_c^{3leg}/2\pi \approx 18.70$ MHz when $\delta = \eta/2$ and $\eta/2\pi=250$ MHz, consistent with our analysis in Sec.~\ref{infiniteresults}.  The latter is within the capabilities of current experiments.

 One could envision alternative approaches, for example directly studying the scaling of $\langle \Delta \hat P^2\rangle$ with system size.  In practice the best way to extract the phase boundary will depend on experimental details.  The one presented here is concrete, and appears to give good results.  The procedure in Sec.~\ref{sec:stateprep} is constructed to produce the ground state, but one might envision alternative approaches which produce a finite temperature ensemble of states \cite{Chen2025,lloyd2026quantumthermalstatepreparation}.  In that case the scaling relationships would take slightly different forms \cite{song2010}. 


\section{State Preparation}\label{sec:stateprep}

The Mott insulating state at $g=0$ is a product state, and hence easy to prepare.  We propose creating that state, then adiabatically ramping  up $g$ to explore the entire phase diagram.  The challenge here is that the timescale for adiabaticity is set by the inverse of the excitation gap $\tau\sim 1/\Delta$ \cite{Chandran2012,Gardas2017}.  Both in the superfluid phase, and at the phase boundary,
$\Delta\sim L^{-1}$, suggesting that the state preparation time should be proportional to the system size. 



We model these dynamics, linearly sweeping the coupling from $g=0$ to $g=g_f$ in time $\tau$, taking $g(t)= g_\text{f } t/\tau$.  We use a finite matrix product state description, using the ``Time Evolving Block Decimation" algorithm to evolve the state \cite{Vidal2003,AJDaley2004}.  We choose our time step $\delta t$ sufficiently small, and our bond dimension $\chi$ sufficiently large that our results are independent of $\delta t$ and $\chi$. We have confirmed that we get the same results when we instead use the ``Time Dependent Variational Principle" algorithm \cite{Haegeman2011,Haegeman2016}.  
As an experimentally accessible measure of the fidelity of state preparation, we calculate the polarization fluctuations $\langle \Delta \hat P^2\rangle$ during the sweep, comparing it to the expectation value in the ground state.  

\begin{figure}
    \centering
    \includegraphics[width=1.05\columnwidth]{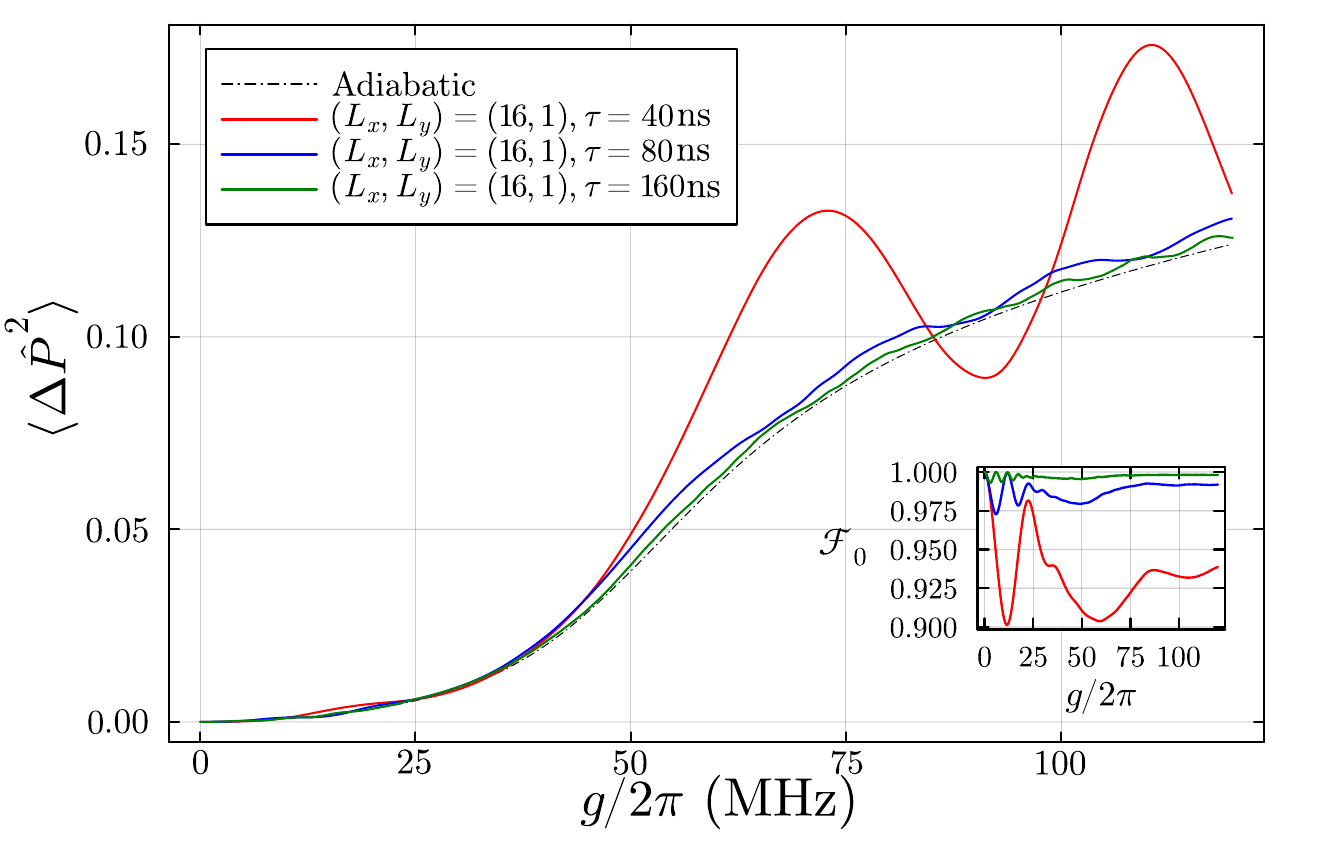}
    \caption{Dynamics of state preparation for a $L_x\times L_y=(16\times 1)$ system  with $\eta/2\pi =$ 0.25GHz and $\delta = \eta/2$.  The coupling constant $g$ is swept linearly from $g_i/2\pi=0 $ MHz to $g_f/2\pi=120 $ MHz in time $\tau=40,80,160$ns.  The main figure shows the polarization fluctuations, $\langle \Delta \hat P^2\rangle$, while the inset shows the square overlap with the instantaneous ground state.  }
    \label{fig:Xfluc1D_Dynamics}
\end{figure}

We first consider the one-dimensional case, and see how the polarization fluctuations vary with the ramp time $\tau$.  Results are shown in Fig.~\ref{fig:Xfluc1D_Dynamics} for $\eta/2\pi=250$MHz and $\delta=\eta/2$.   The calculations from Sec.~\ref{sec:Thermodynamic_limit} and \ref{finite} suggest that in the thermodynamic limit ($L\to\infty$) the phase boundary occurs at $g_c/2\pi \approx 70$ MHz for those parameters.
The black dashed line corresponds to the adiabatic result, while the three colored lines correspond to different sweep rates.  One can see that when $\tau=40$ns the polarization shows  substantial deviations from the adiabatic results.  For $\tau= 80, 160$ns, however, the deviations are quite small.  The inset shows the fidelity, corresponding to the overlap between the instantaneous ground state and the time-evolved wavefunction.  It appears that the polarization fluctuations are a good surrogate for this fidelity.  

In an experiment one could follow this exact procedure, measuring $\langle \Delta \hat P^2\rangle$ for sweeps with several different rates.  If the observed polarization fluctuations are independent of sweep rates, then the results are indicative of the adiabatic limit.  Finite size scaling, as discussed in Sec.~\ref{finite} could then be used to identify the critical point.  One would measure $\langle \Delta \hat P^2\rangle$ for a range of system sizes and values of $g$.  One would then fit the peak of $d\langle \Delta \hat P^2\rangle/dg$ to the expected Kosterlitz-Thouless form. 

\begin{figure}
    
    
        \includegraphics[width=1.0\columnwidth]{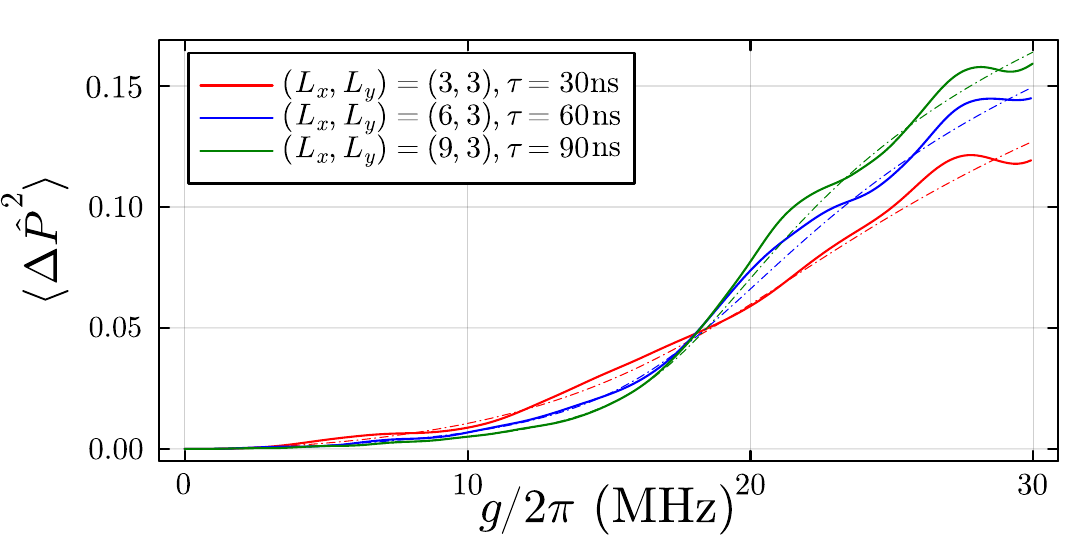}
        \label{fig:State_prep_3leg_slow}

    \medskip

        \includegraphics[width=1.0\columnwidth]{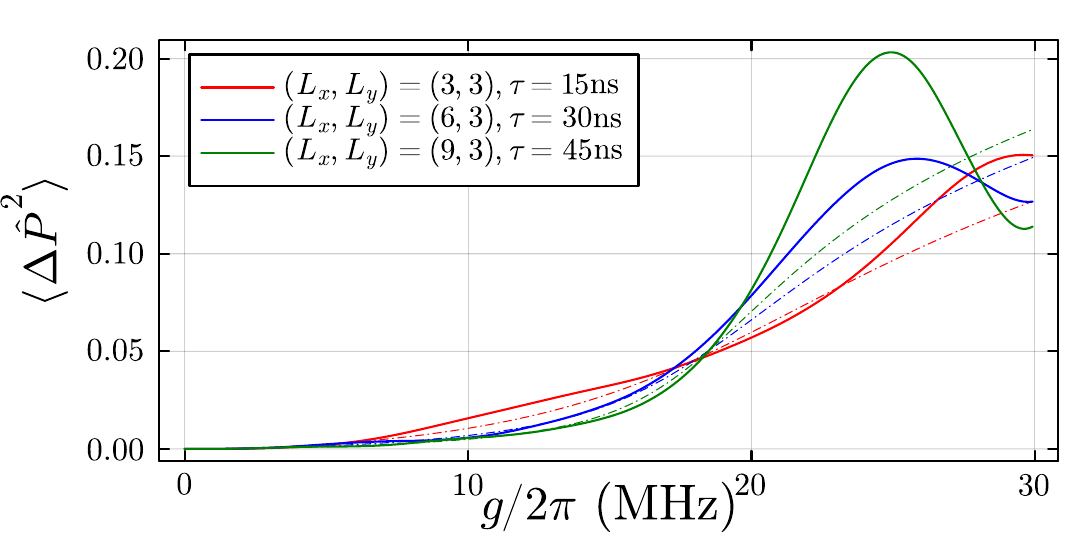}
        \label{fig:State_prep_3leg_fast}

    \caption{Dynamics of state preparation 
    for a $L_x\times L_y=(3,3), (6,3),$ and $(9\times 3)$ ladder with
     $\eta/2\pi =$ 0.25GHz and $\delta = \eta/2$.  The coupling constant $g$ is swept linearly  from $g_i/2\pi = 0 $ MHz to $g_f/2\pi=  30 $ MHz in time $\tau$.  Top:  $\tau=L_x\times 10$ns; Bottom:
     $\tau=L_x\times 5$ns. 
     }
    \label{fig:Xfluc1D_Scaling}
\end{figure}

As already discussed, current experiments are limited to coupling constants $g/2\pi\lesssim 30$ MHz, and it would be challenging to observe the phase transition in Fig.~\ref{fig:Xfluc1D_Dynamics}.  Thus in Fig.~\ref{fig:Xfluc1D_Scaling} we consider the 3-leg ladder, where $g_c/2\pi\approx 19$ MHz when  $\eta/2\pi=250$MHz and $\delta=\eta/2$.  Here  we explore the system size dependence of state preparation, taking  three system sizes, $(L_x,L_y)$ with $L_y=3$ and $L_x=3,6,9$.  In the top panel we take $\tau=30,60,90$ns, so that $\tau\propto 1/L_x$.  All three cases show small deviations from the adiabatic expectation.  Although we do not plot it here, the overlap with the adiabatic state is roughly $99.5\%$.  In the bottom panel, we double the sweep rate.  All three system sizes now yield large deviations.  The overlap with the adiabatic state falls to $95\%$.
These timescales are accessible to current experiments, where 
$T_1$ times are of order  $\mathcal{O}(10\mu\text{s})$ \cite{GoogleQuantumAI_Willow2025,GoogleQuantumAI_ErrorCorrection2025
}. 

\section{Summary}

Our understanding of the physical world is earned through a well established cycle of experiment and model building \cite{Kuhn1970-un}.  
Quantum computers can be an important part of this endeavor,
allowing us to test hypotheses about how the world works.  Depending on the context they can play a number of different roles:  They are sophisticated quantum experiments  which can display novel surprises.  They are also computational tools, which can play a role in solving our models of the world.  Furthermore, these devices act as `quantum wind tunnels,' and are suitable for engineering tasks including prototyping and designing.  

The checkerboard Bose Hubbard model provides the ideal setting for this task of producing artificial quantum matter.  This system displays  superfluid and insulating phases, which in an experiment can be  distinguished by the size dependence of the polarization, defined in Eq.~(\ref{pdef}).  In the insulating state the statistics of the polarization gives one access to the localization length, while in the one dimensional limit of the superfluid state one can extract a Luttinger parameter.

The couplings required to observe these phases, and the transitions between them, are accessible to current experiments.  This can be contrasted with the vanilla Bose Hubbard model where, at commensurate filling, it is challenging to achieve the strong inter-qubit coupling required to observe the superfluid-phase.

For ladders with an odd number of legs, there are two topologically distinct insulating states that cannot be mapped onto one another without either breaking reflection symmetry or entering the superfluid phase.  
This topological restriction does not apply for ladders with an even number of legs. 
In addition to exploring this rich phase diagram, one can use these insulating states as a platform to study charge pumping.

We show how a series of experiments on different system sizes can be used to find the phase transition point using finite-size scaling. Additionally, 
we analyzed adiabatic state preparation, finding that the superfluid state can be produced from the insulator by ramping the qubit coupling 
over a timescale $\tau\sim 100$ns for a $9\times 3$ ladder with typical parameters.  The timescale for adiabaticity scales linearly with the system size.

Throughout we made use of the ITensor library  \cite{itensor,itensor-r0.3}.

\section*{Acknowledgements}

TGK acknowledges support from the Gordon and Betty Moore Foundation under GBMF7392 and from the National Science Foundation under grant PHY-2309135 to the Kavli Institute for Theoretical Physics (KITP).  
PP acknowledges support from Google Quantum AI.
EJM and PP acknowledge support from the National Science Foundation under grant PHY-2409403.

\appendix

\section{Extracting Excitation gap from the Local Hamiltonian gap}\label{gapextrapolation}

In this Appendix we detail how we use the local projected Hamiltonian to approximate the spectral gap, including our implementation of finite entanglement scaling.  We build on arguments from \cite{chepiga2017,chepiga2020,eberharter2023}, where various researchers studied 
how
information about the many-body excitation spectrum 
is encoded in
the eigenvalues of the Hamiltonian projected onto the set of local variations of a given iMPS tensor. 
In particular, references \cite{chepiga2017,chepiga2020} found that the local Hamiltonian spectrum appeared to agree with the exact many-body spectrum in critical, gapless phases. Reference \cite{eberharter2023} applied this idea to compute the characteristic velocity of excitations in a gapless phase. 

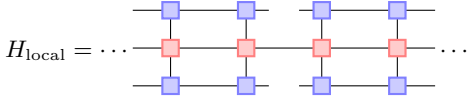
\begin{figure}
\begin{align*}
H_{\rm local} &=
\cdots\raisebox{-0.4\height}{\begin{tikzpicture}
\draw (0.5,1) -- (2.3,1);
\draw (2.7,1) -- (4.5,1);
    \draw (0.5,0.5) -- (4.5,0.5);
    \draw (0.5,0) -- (2.3,0);
\draw (2.7,0) -- (4.5,0);
    \draw (1,1) -- (1,0.);
    \draw (2,1) -- (2,0.);
    \draw (3,1) -- (3,0.);
    \draw (4,1) -- (4,0.);
    \node[tensor] at (1,1) {};
    \node[H] at (1,0.5) {};
    \node[tensor] at (1,0) {};
    \node[tensor] at (2,1) {};
    \node[H] at (2,0.5) {};
    \node[tensor] at (2,0) {};
    \node[tensor] at (3,1) {};
    \node[H] at (3,0.5) {};
    \node[tensor] at (3,0) {};
    \node[tensor] at (4,1) {};
    \node[H] at (4,0.5) {};
    \node[tensor] at (4,0) {};
\end{tikzpicture}}\cdots
\end{align*}
\caption{Diagramatic representation of the local Hamiltonian.  The blue boxes correspond to the matrix product state representation of the wavefunctions, and the red boxes form a matrix product operator representation of the Hamiltonian.  The uncontracted bonds represent external indices.}
\label{fig:localH}
\end{figure}

For our calculations, we consider the Hamiltonian projected onto variations of a single bond-tensor of the iMPS, i.e. a tensor with only virtual indices. 
This local Hamiltonian is shown schematically in Fig.~\ref{fig:localH}. It can be viewed as a $\chi^2\times \chi^2$ matrix, where $\chi$ is the bond dimension of the matrix product state.
Eigenvectors of this local Hamiltonian eigenvalue problem will constitute an orthogonal set of excited states, which are constructed simply by making a local change to the MPS.  These states explicitly break translation invariance.

A valid worry is that excited states of the local Hamiltonian correspond to a very restrictive ansatz for excitations of the system: they are identical to the ground-state wavefunction except for a single bond tensor. 
Conversely, the excitations of our model are typically 
delocalized over the entire system.
In the limit of large bond dimensions, however, a local insertion of this form can have long-range effects that are only cut off by the iMPS correlation length, $\xi_\chi$. If the excitations are dispersive, then the local excitations found by this method can be understood as superpositions of the true excited states confined to some region -- in effect, a particle-in-a-box. 

This intuitive picture allows for a natural means for extrapolating the local excitation energy as the bond dimension is increased. Let us construct an analogy between local Hamiltonian excitations and those of a particle trapped in a finite well. At finite bond dimension, $\chi$, the iMPS will display some finite correlation length, $\xi_\chi$, as well as some finite local Hamiltonian gap, $\Delta_\chi$. We'll also consider higher-energy eigenstates of the local Hamiltonian, $\Delta^{(n=2,3,\ldots)}_\chi$, where $\Delta_\chi=\Delta_\chi^{(1)}$. In the limit $\chi\to\infty$, one will recover the asymptotic correlation length, $\xi$, and energy gap, $\Delta$:
\begin{eqnarray}
    \lim_{\chi\to\infty}\xi_\chi&=&\xi \\
    \lim_{\chi\to\infty}\Delta_\chi&=&\Delta.
\end{eqnarray}
In order to extrapolate $\Delta_\chi$, however, we must know how it approaches $\Delta$ for large $\chi$. Let us now posit that the true excitations are dispersive, and hence are eigenstates of the momentum operator. Their energies can hence be written as $E_k=\Delta+\epsilon_k$ where ${\rm min}_k\epsilon_k=0$. The local Hamiltonian eigenstates are simply superpositions of these momentum states confined to a particular region in space. Hence, we may model their energies as those of a particle-in-a-box: $E^{(n=1,2,\ldots)}=\Delta+\frac{h^2n^2}{2mL}$ for a box of length $L$ and a particle of mass $m$. The effective mass is determined by the curvature of the dispersion, $\epsilon_k$, in the vicinity of its minimum. The length of the ``box", in this analogy, is the parameter that controls how $\Delta_\chi$ approaches $\Delta$ as $\chi\to\infty$. Rather than attempting to determine these bond-dimension-dependent parameters $L$ and $m$, however,
we note that the form of $E^{(n)}$ implies a convenient relationship between the iMPS local gap, $\Delta_\chi$, and the spacing between higher-energy eigenvalues:
\begin{equation}
    \Delta^{(1)}_\chi-\Delta\propto \Delta^{(n)}_\chi-\Delta^{(n-1)}_\chi.
\end{equation}
Note of course that the proportionality constant will be $n$-dependent, but the $\chi$-dependent factors cancel. Hence, one may extrapolate the local Hamiltonian eigenvalues without reference to this undetermined length scale, $L$. This particle-in-a-box form can only apply for the first few low-energy excitations of the local Hamiltonian, so it is advantageous to choose $n$ to be small. In Fig.~\ref{fig:nleg_gaps}, we determine the asymptotic spectral gaps using a linear least-squares fit of the form $\Delta^{(1)}_\chi=\Delta+a(\Delta^{(2)}_\chi-\Delta^{(1)}_\chi)$ where $\Delta$ and $a$ are fitting parameters. We find that the linear fit applies well to our data for $\chi\gtrsim50$.

This procedure for extrapolating the local Hamiltonian eigenvalues bears resemblance to the procedure used in Ref.~\cite{rams2018} (and subsequently in Ref.~\cite{vanhecke2019}) to extrapolate finite-$\chi$ correlation lengths as $\chi\to\infty$. The relationship between the matrix of correlation lengths and the local Hamiltonian eigenvalues was explored for gapless systems in Ref.~\cite{eberharter2023}. Our work suggests that a more robust low-energy correspondence may apply even for gapped systems, particularly when the bond dimension is large~\cite{white2026sitebasisexcitationansatz}.

\section{Center of Mass fluctuations in the Superfluid Phase\label{sec:ll}}
\subsection{Luttinger Liquid with Boundaries}\label{ll_bg}
Here we calculate the polarizaton fluctuations of a finite Luttinger liquid, which is the continuum limit of our model of the superfluid phase.  
 The low energy excitations are described by a Hamiltonian~\cite{giamarchi}
\begin{equation}\label{ll}
    \hat{H}=\frac{u}{2\pi}\int_{0}^{L}dx[K(\pi\hat{\Pi})^2+\frac{1}{K}(\partial_x \hat{\varphi})^2]
\end{equation}
Here $\hat{\varphi}(x)$ is related to the particle density $\hat \rho(x)=\sum_j \hat n_j \delta(x-x_j)$ by
\begin{align}
\hat \rho(x) &=\bar\rho+\frac{1}{\pi}\partial_x \hat{\varphi}(x),
\end{align}
where $\bar\rho=\bar n/a$ is the average density, $\bar n$ is the average site occupation, and $a$ is the lattice spacing.
The conjugate variable,
 $\hat{\Pi}(x)$ is related to the momentum density, and these operators obey  commutation relations $\left[ \hat\Pi(x),\hat\varphi(y) \right] = i\delta(x-y)$.
The speed of sound, $u$, and the Luttinger parameter, $K$, encode the mechanical properties the fluid.  Larger $K$ corresponds to a more compliant system with larger density fluctuations.  The ground state of Eq.~(\ref{ll}) has algebraically decaying correlation functions whose power laws are controlled by $K$. 

The connection between $K$ and microscopic parameters are non-trivial.    The hard core Bose gas at incommensurate filling maps onto a non-interacting Fermi gas with $K=1$.  Weaker interactions lead to larger $K$, and $K\to\infty$ for the weakly interacting Bose gas.  At the superfluid-insulator phase transition $K$ takes on a universal value, $K=2$.


The polarization fluctuations are then 
\begin{align} \label{denscorr_ll}
\langle \Delta \hat P^2\rangle=
&\frac{1}{L_x^2}\int_{0}^{L_x} \!\!dx dx^\prime x x' \langle(\hat{\rho}(x)-\bar \rho)(\hat{\rho}(x')-\bar \rho)\rangle\nonumber\\
&= \frac{1}{\pi^2 L_x^2}\int_{0}^{L_x} \!\!dx dx^\prime\, \langle
\hat{\varphi}^\prime(x) \hat{\varphi}^\prime(x^\prime)
\rangle.
\end{align}
We perform a normal-mode expansion of the Hamiltonian,
\begin{align} \label{ll_modes}
    \hat{\varphi}(x)&=\sum_{n=1}^{N}\hat{\varphi}_nu_n(x)& \hat{\Pi}(x) &=\sum_{i=n}^{N}\hat{\Pi}_nu_n(x)
\end{align}
%
where the mode wavefunctions $u_n(x) = \sqrt{\frac{2}{L}}\sin(\frac{\pi n x}{L})$ are chosen by requiring that currents vanish at the boundaries $x=0,L$. Here $[\hat \Pi_n,\hat \varphi_m]=i\delta_{nm}$ correspond to the canonical variables of a set of decoupled harmonic oscillators that represent the low energy excitations.  In terms of these variables, 


\begin{equation}\label{corrected}
\begin{split}
    \hat{H}&=\frac{u\pi}{L}\left(\frac{ KL}{2}\sum_{n}^{}\hat\Pi_n^2+\frac{1}{2 KL}\sum_{n}^{}n^2\hat\varphi_n^2\right)
\end{split}
\end{equation}
In the ground state
$
\langle \hat \varphi_n \hat\varphi_m\rangle 
=\delta_{nm}(KL)/(2n)
$
 Inserting Eq.~(\ref{ll_modes}) into Eq.~(\ref{denscorr_ll}) and evaluating the ground-state expectation values accordingly, we arrive at
\begin{align}
    \langle\Delta \hat{P}^2\rangle &= \frac{4K}{\pi^4}\sum_{n\in \text{odd}}^{L_x}\frac{1}{n^3}
\xrightarrow[L_x\to\infty]{} \frac{7\zeta(3)}{2\pi^4}K
\label{polresult}
\end{align}
where we have used the identity
$\sum_{n\in\text{odd}}\frac{1}{n^s}=\left(1-\frac{1}{2^s}\right)\zeta(s)$, where $\zeta(s)$ is the Riemann zeta function.

\subsection{Free Fermion Calculation}

The Luttinger liquid description in Appendix~\ref{ll_bg} involves a long-wavelength approximation and expands about a uniform density.  Here we calculate the polarization fluctuations in the case of hard-core bosons or non-interacting fermions, without making those assumptions.  
The result agrees with the expression from Appendix~\ref{ll_bg} with $K=1$.  This calculation lends further confidence to the validity of those results.

Equation~(\ref{polresult}) is independent of density, and  we consider the case where there are $N$ nonininteracting fermions on $L-1$ sites, forming a Slater determinant filling orbitals
$\psi^\alpha_j=\sqrt{\frac{2}{L}}\sin\frac{\pi \alpha j}{L}$ with $\alpha=1,2,\cdots N$, while the position $j$ takes on values $j=1,2,\cdots L-1$.  We take $N,L\to \infty$ with  $n=N/(L-1)\approx N/L$ fixed.  


Our goal is to compute the sum:
\begin{align}
    \langle \Delta \hat P^2\rangle&=\frac{1}{L^2}\sum_{i,j}ij\langle (\hat n_i-n)(\hat n_j- n)\rangle.
\end{align}
A straightforward application of Wick's theorem allows us to write the polarization fluctuations as a sum over particle and hole expectation values
\begin{align}
    \langle \Delta \hat P^2\rangle&=\frac{1}{L^2}\sum_{ij} i j \langle \hat \psi_i^\dagger \hat \psi_j\rangle\langle \hat \psi_i \hat \psi_j^\dagger\rangle.
    \label{polsum}
\end{align}
We introduce the dipole matrix elements
\begin{align}\label{overlaps}
F_{\alpha\beta}&= \frac{1}{L} \sum_j j \psi_j^\alpha \psi_j^\beta.
\end{align}
which can be used to express
$\langle \Delta \hat P^2\rangle$
as a sum over occupied and unoccupied modes,
\begin{align}
   \langle \Delta \hat P^2\rangle &=
   \sum_{
   \substack{\alpha\leq N\\\beta>N} }
   F_{\alpha\beta}^2.
\end{align}  
By symmetry the spatial sums in Eq.~(\ref{overlaps}) vanish unless $\alpha$ and $\beta$ have opposite parity, where one finds
\begin{align}
F_{\alpha\beta} &= \frac{1}{2L^2} \left(
\frac{1}{\sin^2 \frac{\pi}{2L}(\beta-\alpha)}-
\frac{1}{\sin^2 \frac{\pi}{2L}(\beta+\alpha)}
\right)
\end{align}
Given this structure, the sum in Eq.~(\ref{polsum}) will be dominated by contributions from the Fermi surface, where $\alpha$ and $\beta$ are near $N$.  For these we can 
approximate
\begin{align}
F_{\alpha\beta}
&\approx\frac{2}{\pi^2}\frac{1}{(\beta-\alpha)^2}.
\end{align}
There are $n_\gamma=\gamma$ ways of choosing $\alpha$ and $\beta$ so that  $\gamma=\beta-\alpha$.  We can therefore write
\begin{align}
\langle \Delta \hat P^2\rangle &=\sum_{\gamma\ \in \text{odd}} \left(\frac{2}{\pi^2} \frac{1}{\gamma^2}\right)^2 n_\gamma.
\end{align}
This is the same sum that appears in Eq.~(\ref{polresult}), with $K=1$, and hence we find the expected result that $\langle \Delta \hat P^2\rangle=7\zeta(3)/(2\pi^4)$.

\bibliographystyle{apsrev4-2}
\bibliography{bhm.bib}




\end{document}